\documentclass[%
aps,
twocolumn, prd,
showpacs,preprintnumbers,nofootinbib,
amsmath,amssymb,
aps,
floatfix,
10pt,
superscriptaddress]{revtex4-1}

\RequirePackage{lineno}
\usepackage{float}
\usepackage{graphicx}
\usepackage{dcolumn}
\usepackage{bm}
\usepackage{color}
\usepackage{array}
\usepackage{multirow}
\usepackage[normalem]{ulem}
\usepackage{epstopdf}
\usepackage{ulem}
\usepackage{natbib}
\usepackage{mathtools}
\usepackage{braket}
\usepackage{gensymb}

\usepackage{hyperref}
\hypersetup{
    colorlinks=true,
    linkcolor=blue,
    filecolor=magenta,
    urlcolor=cyan,
}

\begin{document}

\title{NuRIA: Numerical Relativity Injection Analysis of signals from generically spinning intermediate mass black hole binaries in Advanced LIGO data}

\author{Koustav~Chandra}
\email{koustav.chandra@iitb.ac.in}
\affiliation{Department of Physics, Indian Institute of Technology Bombay, Powai, Mumbai, Maharashtra 400076, India}
\author{V. Gayathri}
\email{gayathri.v@ligo.org }
\affiliation{Department of Physics, Indian Institute of Technology Bombay, Powai, Mumbai, Maharashtra 400076, India}
\author{Juan Calder\'on~Bustillo}
\email{bustillo@phy.cuhk.edu.hk}
\affiliation{Monash Centre for Astrophysics, School of Physics and Astronomy, Monash University, VIC 3800, Australia}
\affiliation{OzGrav: The ARC Centre of Excellence for Gravitational-Wave Discovery, Clayton, VIC 3800, Australia}
\affiliation{Department of Physics, The Chinese University of Hong Kong, Shatin, N.T., Hong Kong}

\author{Archana~Pai}
\email{archana@phy.iitb.ac.in}
\affiliation{Department of Physics, Indian Institute of Technology Bombay, Powai, Mumbai, Maharashtra 400076, India}

\begin{abstract}
The advent of gravitational wave (GW) astronomy has provided us with observations of black holes more massive than those known from X-ray astronomy. However, the observation of an intermediate-mass black hole (IMBH) remains a big challenge. After their second observing run, the LIGO \& Virgo Scientific collaborations (LVC) placed upper limits on the coalescence rate density of non-precessing IMBH binaries (IMBHBs). In this Numerical Relativity Injection Analysis (NuRIA), we explore the sensitivity of two of the search pipelines used by the LVC to signals from 69 numerically simulated IMBHBs with total mass greater than $200 M_\odot$ having generic spins, out of which 27 have a precessing orbital plane. In particular, we compare the matched-filter search PyCBC, and the coherent model-independent search technique cWB. We find that, in general, cWB is more sensitive to IMBHBs than PyCBC, with the difference in sensitivity depending on the masses and spins of the source. Consequently, we use cWB to place the first upper limits on the merger rate of generically spinning IMBH binaries using publicly available data from the first Advanced LIGO observing run.
\end{abstract}
\maketitle
\section{Introduction}
During its first two observing runs (respectively O1 and O2), the gravitational-wave detector network formed by Advanced LIGO~\cite{aLIGODetector}  and Virgo~\cite{TheVirgo:2014hva, Acernese_2018} detected the coalescence of one binary neutron star~\cite{GW170817-DETECTION}) and ten binary black holes (BBHs) ~\cite{GW150914-DETECTION, GW151226-DETECTION, GW170104-DETECTION, GW170608-DETECTION, GW170817-DETECTION, GW170814-DETECTION, TheLIGOScientific:2016pea, O1_O2_CATALOG}. Also, several independent analyses have reported additional candidates based on their study on publicly available data ~\cite{1-OGC,2-OGC, IAS-Catalogue, IAS-Catalogue_II, Antelis:2018smo}

Since the beginning of the third observing run with much-improved sensitivity, the network has been reporting alerts for astrophysical signals every week, and $\sim {\cal{O}}(100)$ detections are expected by the end of the run. Not only will these observations allow us to study the population and properties of compact binary objects but also lead to the observation of new unobserved sources. For example, the most massive binary neutron star
system till date is detected recently in the third observing run \cite{GW190425-Discovery}.

In this Numerical Relativity Injection Analysis (NuRIA), we focus on yet another elusive source: intermediate-mass black holes (IMBHs)~\cite{EardleyPress1975,Rees1978,Bahcall1975,Begelman1978,Quinlan1990,Greene_Review}. These are usually defined as black holes (BHs) with masses in the range of $10^2-10^5 M_\odot$. They are the missing link between the stellar-mass black holes (SBHs) observed so far by GW detectors (roughly in $18 M_\odot $ to $85 M_\odot$~\cite{O1_O2_CATALOG}) and the supermassive black holes (SMBHs) with masses larger than $10^5 M_\odot$ that are known to lay in the centres of most galaxies. Despite several indirect pieces of evidence for the existence of these objects from electromagnetic measurements, there is no conclusive direct observation. Such observation would set a milestone for astrophysics, shedding light on how a population of SBHs can transition to SMBHs through, for instance, a hierarchical merger channel~\cite{Merzcua:2017,2017mbhe.confE..51K}.

Intermediate mass black hole binaries (IMBHBs) are potentially the loudest sources for current ground-based GW detectors. Despite this, a dedicated search on O1-O2 data reported no detection of any IMBHBs and placed upper limits on their merger rate density~\cite{O1O2IMBHBPaper}. In particular, the most stringent rate upper limit of $0.2 Gpc^{-3} yr^{-1}$ was placed for the case of equal-mass binaries with individual masses $m_1=m_2=100 M_\odot$ and equal aligned spin parameters of $\chi_{1z}=\chi_{2z}=0.8$. \footnote{${\bf \chi_i} = \frac{c\vec{S}_i}{Gm_i^2}$, with $m_i$ and ${\bf S_i}$ being respectively the masses and spins of the two-component objects.}

To place this upper limit, simulated IMBHB signals were injected in the detector data and then recovered with the search algorithms. In \cite{O1O2IMBHBPaper}, for the first time, numerically simulated signals containing all the physics of the IMBHB systems were used, but they were restricted to the systems with BH spins aligned to the orbital plane of the binary. This is a reasonable choice, as the effects of a precessing orbital plane may not be, in principle, observable for short-lived IMBHB signals, vastly dominated by the merger and ringdown emission. However, studies have shown that the effect of precession can be observed in IMBHB systems \cite{Mapelli:2016vca, HeadON}
and hence is interesting to probe the precession effect as detectors become more sensitive to IMBHB systems with an increase in sensitivity.

Hierarchical mergers of BHs in the dense globular clusters are one of the birth-places for IMBHBs. Studies have shown that in such a dense environment, BHs do not carry any preferential spin orientation \cite{Rodriguez:2016kxx}. As a result, binaries formed from these BHs are expected to distribute isotropically in spin orientation resulting in spin-orbit precession.

In this paper, we evaluate the sensitivity of current search algorithms to sources with generic spins and place the first-ever upper limits on their coalescence rate. We use two searches used by the LVC in~\cite{O1O2IMBHBPaper}: the matched-filter algorithm for aligned-spin sources PyCBC~\cite{Canton:2014ena,Usman:2015kfa,pycbc-github} and the unmodelled time-frequency map-based algorithm, coherent WaveBurst (cWB)~\cite{Klimenko:2015ypf}. Consistent with a previous work~\cite{CalderonBustillo:2017skv}, we find cWB is more sensitive than PyCBC to signals from IMBHBs, which strongly deviate from the ``chirp" which PyCBC targets \footnote{As we will describe later, the PyCBC search is currently restricted to ``quadrupole" or $(\ell,m)=(2,\pm 2)$ modes of aligned spin BBHs, omitting higher emission modes.}. Finally, we use cWB to place upper limits on the coalescence rate of a family of IMBHBs with generic spins, using publicly available data from the first Advanced LIGO observing run. We place the most constraining upper limit at $0.36/Gpc^3/yr$ on the merger rate density of precessing equal-mass IMBHs with total masses of $210M_\odot$, improving by a factor of $\sim 3$ on the LVC limits from the O1 run.

\begin{figure*}
    \includegraphics[width= 2\columnwidth]{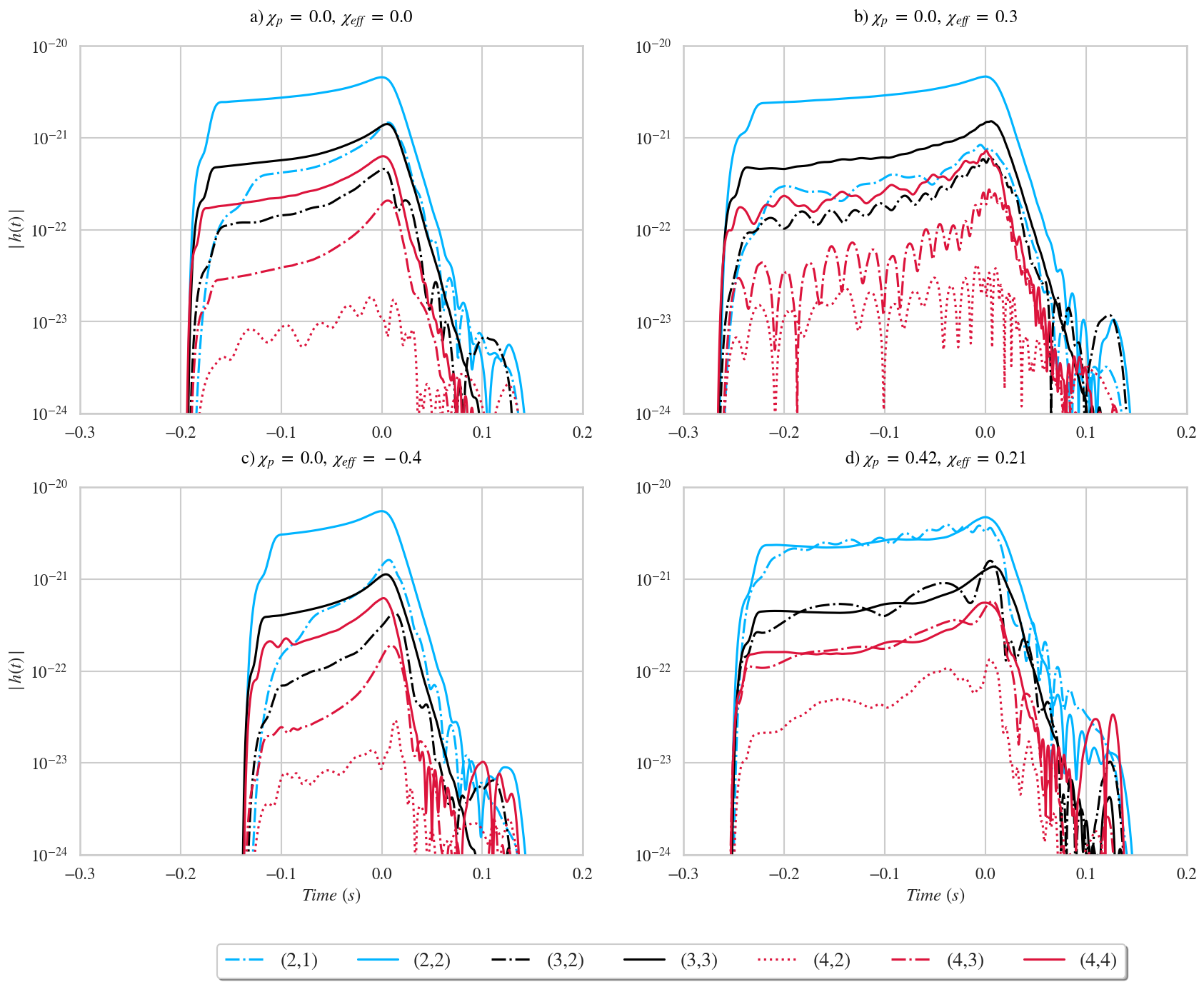}
    \caption{Amplitude modulus of the gravitational-wave modes of different binary black hole systems with mass ratio $q=3$ and varying spins and a total mass of $M=210M_\odot$. The systems have edge-on orientation and are located at a distance of $100$ Mpc. The source parameters are defined at a starting frequency of 16 Hz. We note that the amplitude of the (3,3) and (4,4) mode increases relative to (2,2) mode from no spin case to systems with spins. The spins also impact both the duration of the signal. The strains have been obtained from to the publicly available Georgia Tech catalogue. The corresponding simulations have ID  GT0453, GT0596, GT0838 and GT0874. The oscillations in the (4,2) mode in the top left panel are due to numerical noise. We note, however, that the overlap with an equivalent waveform from the SXS code is around 0.94 and the mode itself is so weak that it does not have any significant impact.}
    \label{fig:modes}
\end{figure*}

The rest of the paper is structured as follows. Section \ref{HM} briefly summarises the impact of precession and higher order-modes on IMBHB signals. Section \ref{Search Algorithm} describes the two search algorithm used in this paper. Section \ref{Search Methodology} describes the analysis setup, including the injections we make in publicly available O1 data~\cite{OpenData} and the evaluation of the sensitivity of the searches. In section \ref{Results}, we first compare the sensitivity of the two searches and then report upper limits on a family of IMBHBs with generic spins. Finally, in section \ref{Conclusion}, we summarise our results.

\begin{figure*}

\includegraphics[width =2\columnwidth]{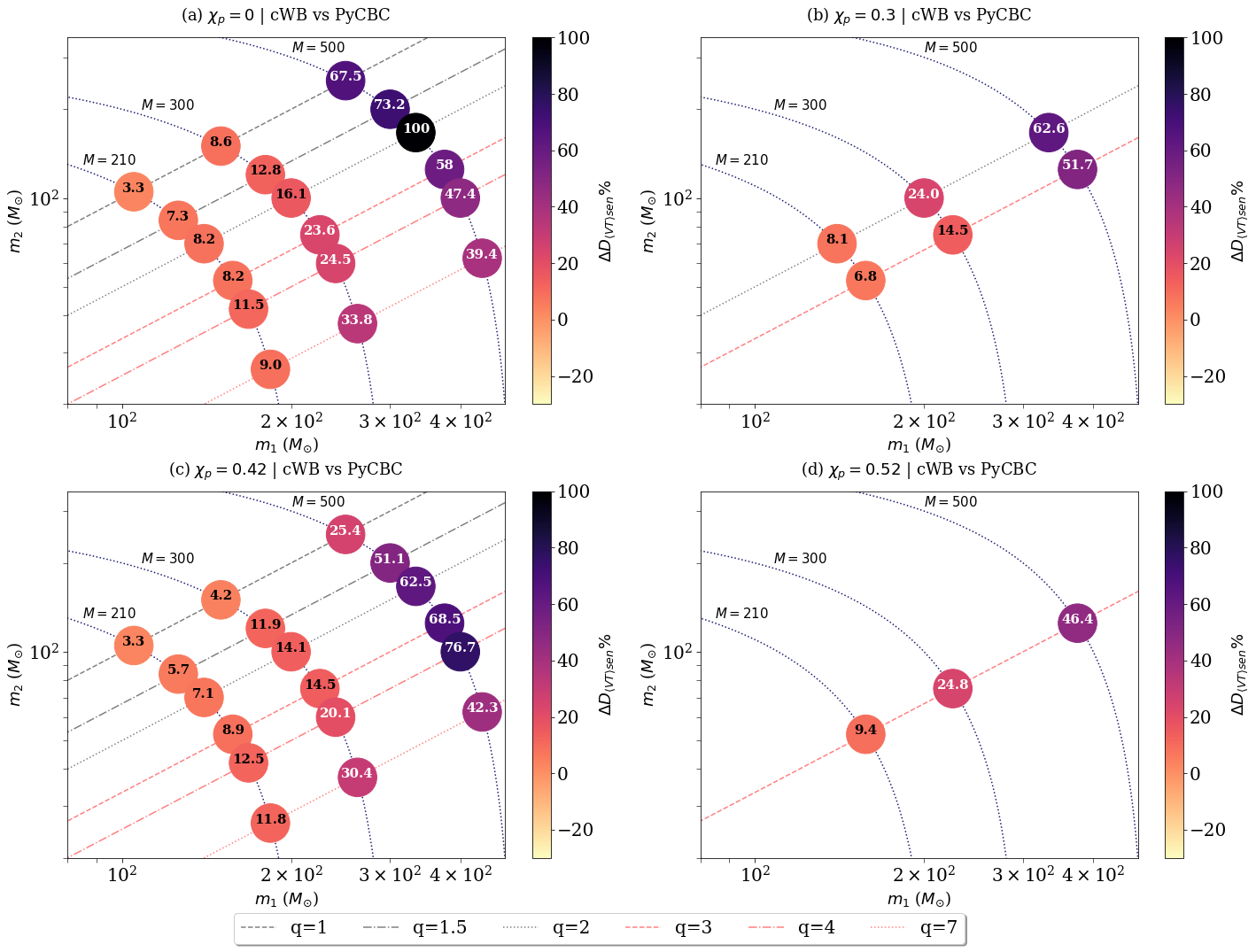}
\caption{cWB vs PyCBC: Comparison of the sensitive distance reach of cWB and PyCBC at IFAR=2.94 yr. The results correspond to injections made in the first two chunks of data from the first Advanced LIGO observing run. The effect of higher modes and precession on the search causes the unmodelled algorithm to outperform the matched-filter search as indicated by the positive values.}
\label{fig:Comparison between the two pipelines}
\end{figure*}

\begin{figure*}

\includegraphics[width =2\columnwidth]{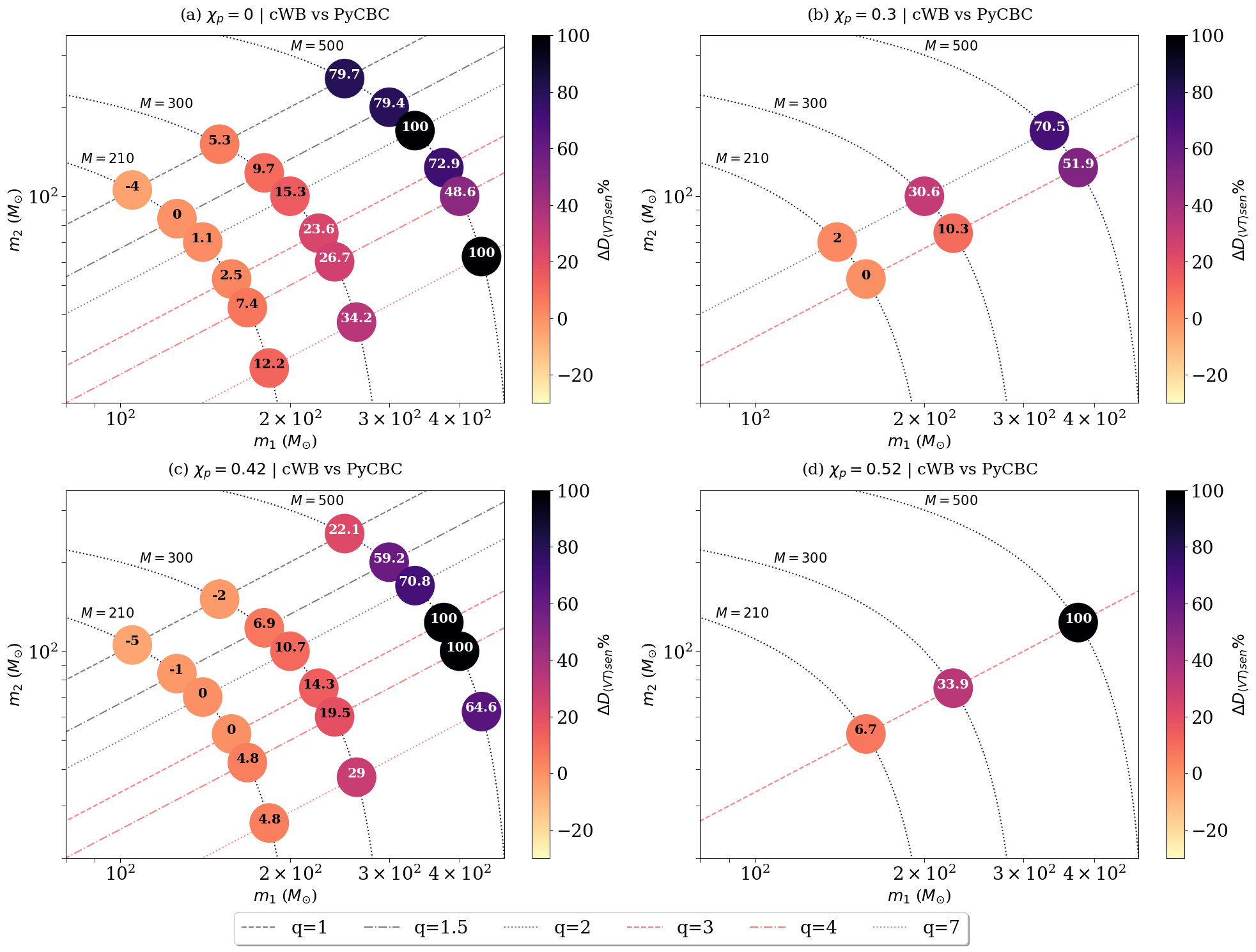}
\caption{cWB vs PyCBC: Comparison of sensitive distance reach of cWB and PyCBC, for IFAR=300yr, as a function of the source parameters. The effect of higher modes and precession causes the unmodelled algorithm to outperform (as indicated by positive values) the matched-filter search, which does not incorporate these two effects. We have checked that the apparent alteration of trends at high mass in panels (a) and (c) is consistent with statistical uncertainty. 
}
\label{fig:Comparison between the two pipelines at IFAR 300}
\end{figure*}

\begin{figure*}
    \includegraphics[width = 2\columnwidth]{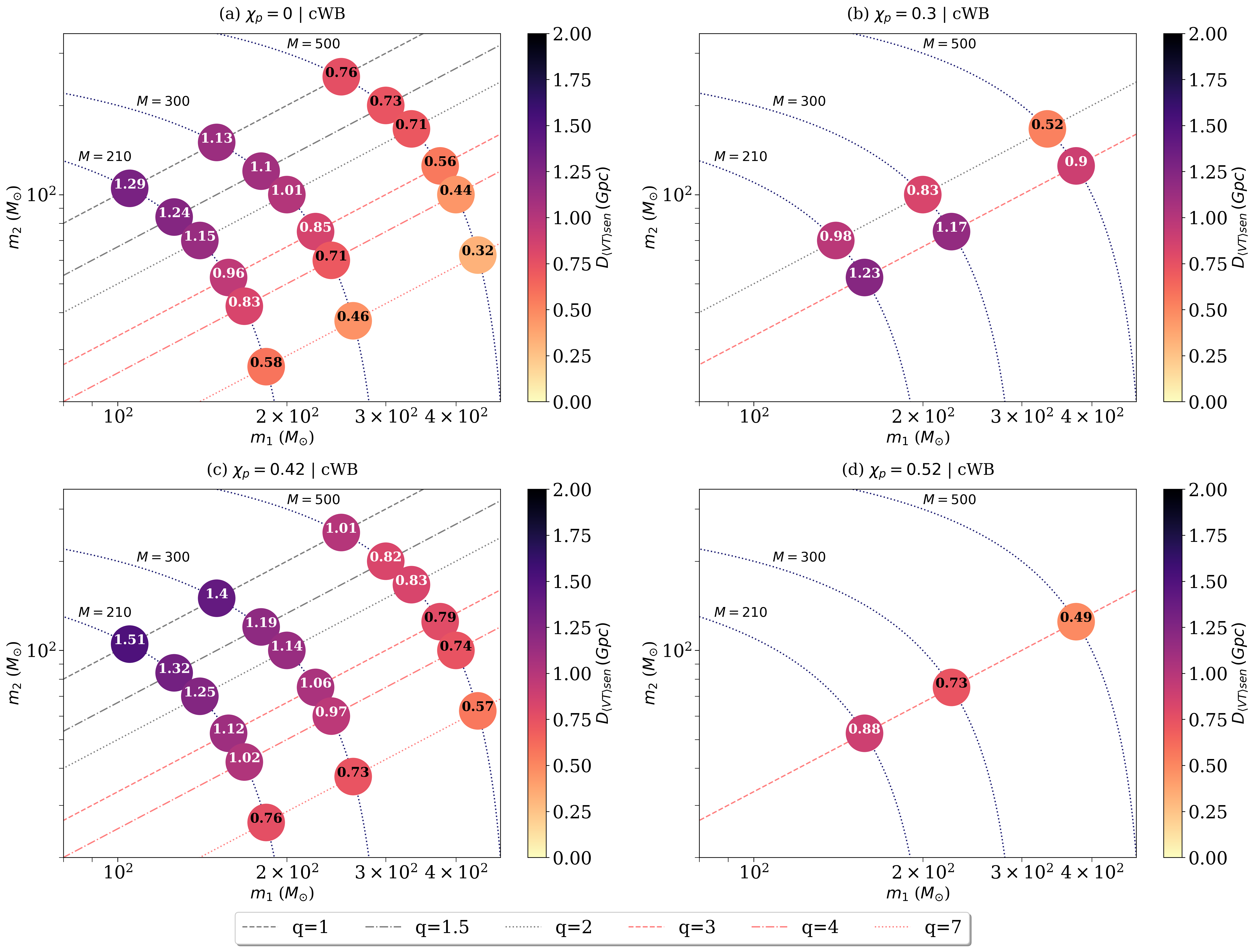}
    \caption{cWB: The plot shows the sensitive distance reach for this set of injections for IFAR = 2.94yr. As a general trend, the sensitivity drops with increasing total mass and mass ratio.}
    \label{fig:cWB}
\end{figure*}

\begin{figure*}
\includegraphics[width=2\columnwidth]{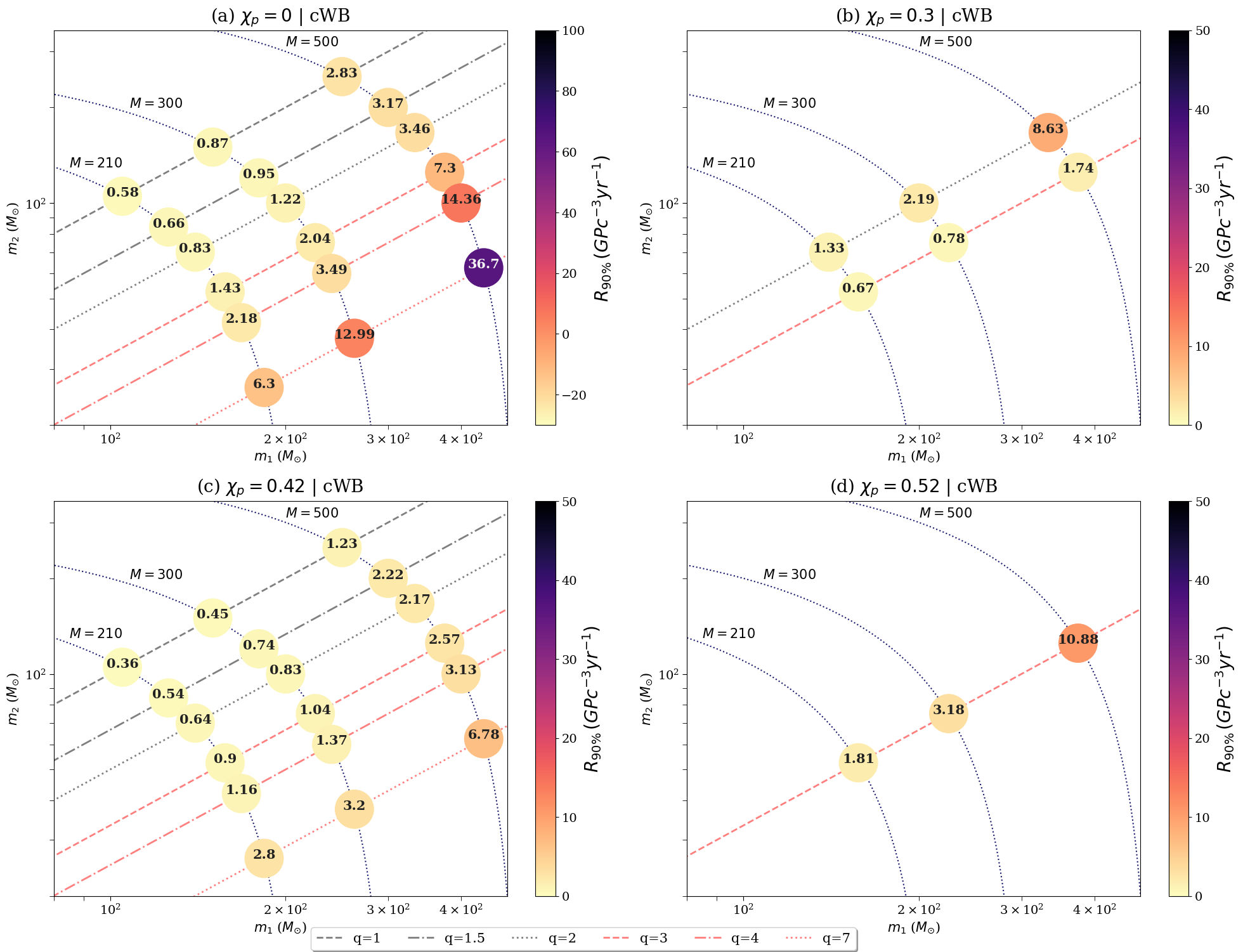}
\caption{90 \% upper limit on merger rate density ($R_{90 \%}$) for our target IMBHB sources, expressed $Gpc^{-3}yr^{-1}$. We set our most stringent upper limit at $0.36 Gpc^{-3}yr^{-1}$ for the symmetric mass system with total mass of 210 $M_\odot$ and $\chi_p=0.42$ and $\chi_{eff}=0.5114$.}
\label{fig:Rate_cWB}
\end{figure*}

\begin{figure*}
    \includegraphics[width = 0.8\textwidth]{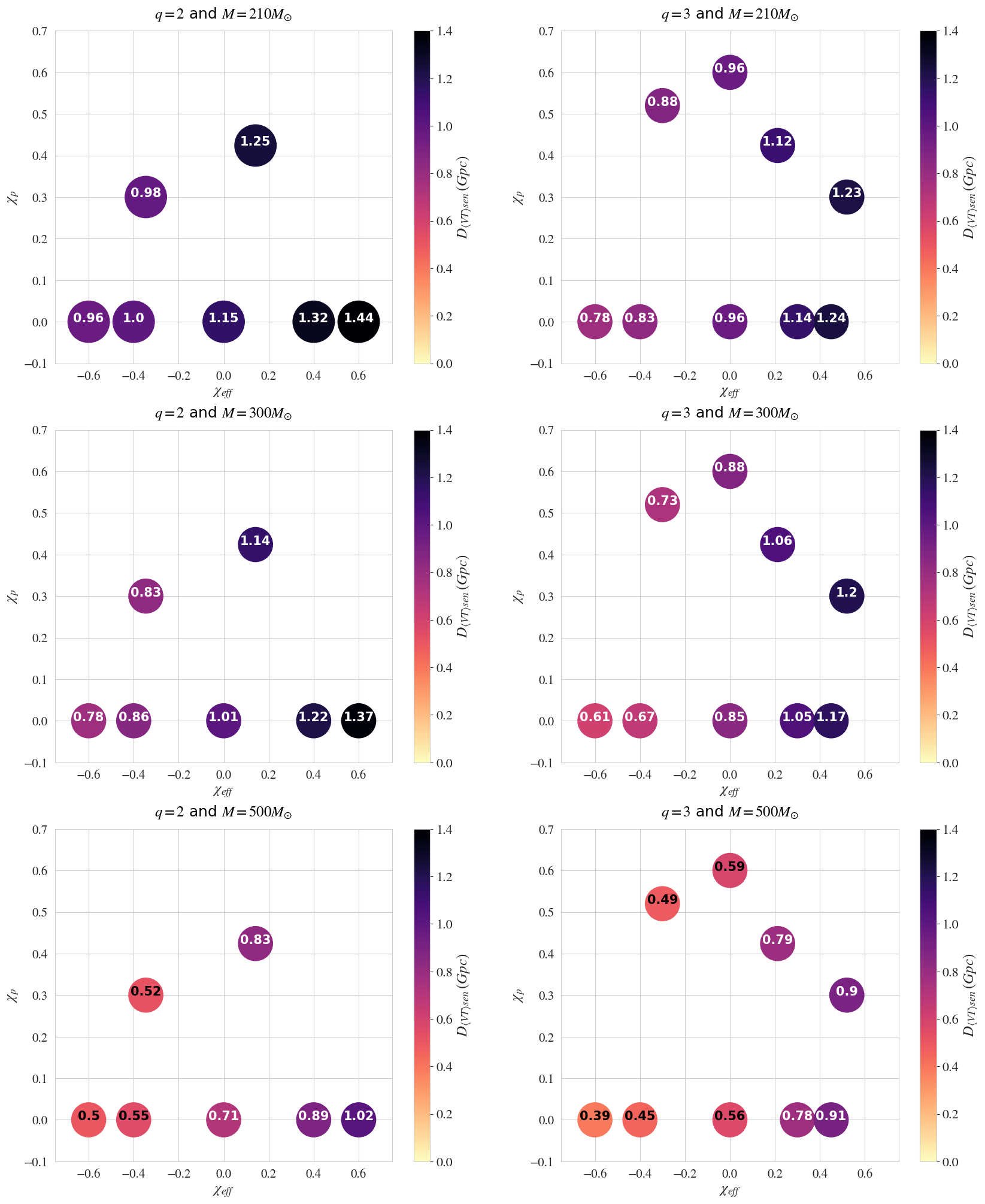}
    \caption{Sensitivity of cWB as a function of the binary spins. When $\chi_p = 0$, an increase in $\chi_{eff}$, causes an increment of the senstive distance reach while a decrease in $\chi_{eff}$ causes the sensitivity to drop.}
    \label{fig:generic spin}
\end{figure*}

\begin{figure*}
    \includegraphics[width=2\columnwidth]{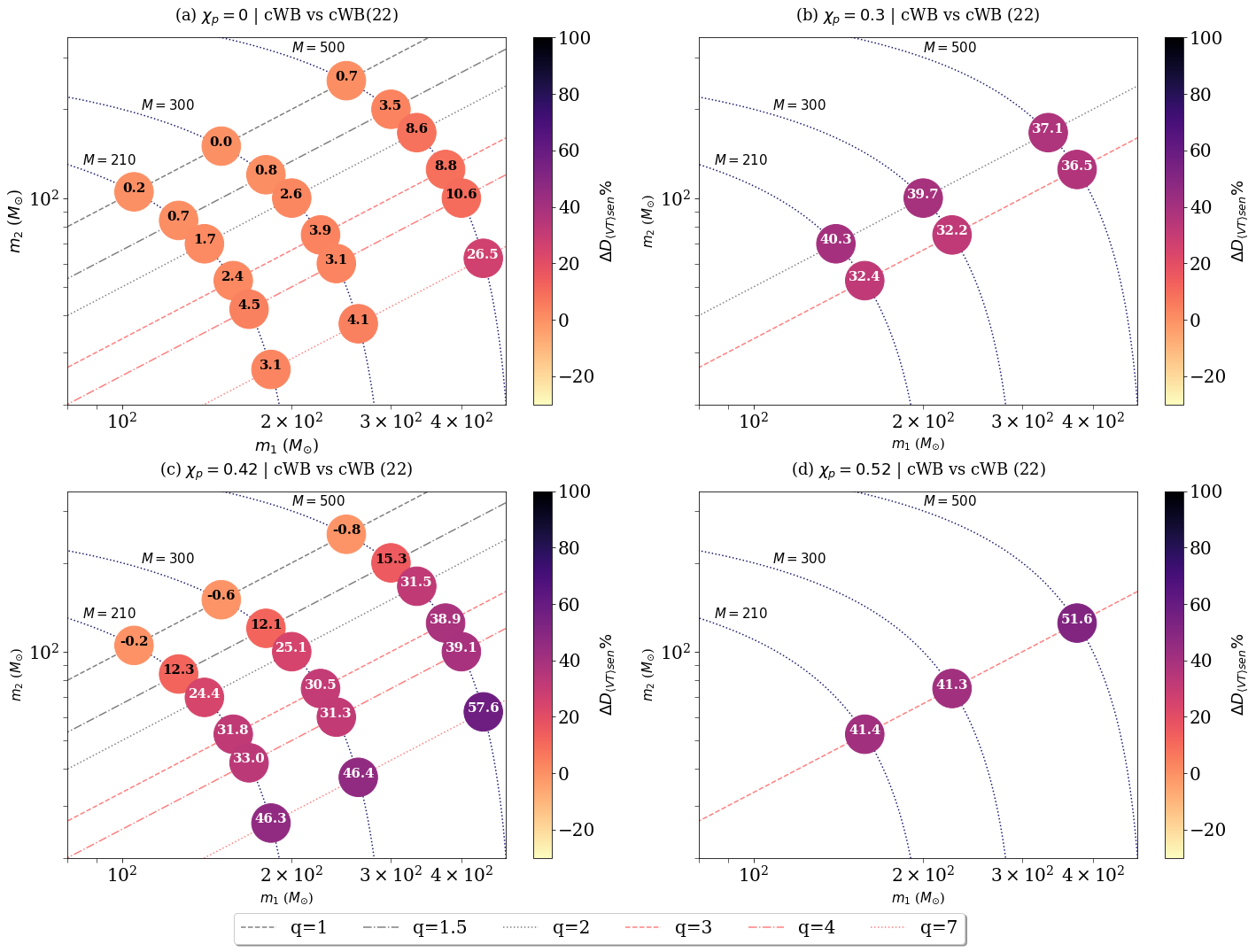}
    \caption{Ratio of the sensitivity of cWB to signals including and omitting higher modes. Values of $\Delta_D$ indicate that cWB is equally sensitive to both kinds of signals while large values indicate an increase in the sensitivity when higher modes are included. The higher modes provide extra signal power as the total mass and mass ratio increase, leading to an important increment of the sensitivity of the search.}
    \label{fig:(2,2) mode vs HM}
\end{figure*}

\section{Source properties and waveform morphology}
\label{HM}

All confirmed gravitational-wave observations of BBHs show a very characteristic chirp morphology. This consists of a monotonically increasing frequency and amplitude during the inspiral and merger stages of the binary, followed by a damped sinusoid with a constant frequency signal during the ringdown. While this is the most extended and studied signal, it is only valid for face-on or nearly face-on BBHs with similar component masses and non-precessing orbital planes. The LIGO BBHs detected so far are consistent with this in terms of their parameters and signal features and henceforth are referred to as {\it vanilla} BBH.

Generically, the two GR modes of polarisation $h_{+,\times}$ of a GW from a BBH are expressed as a superposition of GW modes $h_{l,m}$ weighted by spin-2 spherical harmonics ($Y^{-2}_{l,m}$) as:~\cite{Goldberg:1966uu,maggiore2008gravitational,GW-BOOK-Anderson}:
\begin{equation}\label{eq:spherical}
\begin{aligned}
        h & = h_+ - i h_\times = \sum_{l\geq 2}\sum_{m=-l}^{m=l} Y^{-2}_{l,m}(\iota,\Phi)h_{l,m}(\Xi,D_L;t-t_c)
\end{aligned}
\end{equation}
where the masses and spins of the individual BHs are collectively denoted by $\Xi$. The $(\iota,\Phi)$ are the BBH orientation parameters, $D_L$ is the luminosity distance and $t_c$ denotes the time of coalescence.

For the case of a \textit{vanilla} BBH, the above sum is mostly dominated by the $(2,2)$ quadrupolar mode which is responsible for the characteristic chirp. However, for asymmetric high-mass sources with orbital plane inclinations away from face-on (i.e., towards $\iota=\pi/2$) configuration, higher modes are significant. This does not only lead to more non-trivial waveform morphologies but can also significantly impact the signal loudness ~\cite{Pekowsky:2012sr, Varma:2014jxa,Bustillo:2015qty, Varma:2016dnf,Bustillo:2016gid, Graff:2015bba,CalderonBustillo:2018zuq,CalderonBustillo:2019wwe}.

In addition, spin-induced precession introduces amplitude and phase modulations of the individual modes. Spin-precession is triggered by in-plane BH spin components $\chi_{1,2}^{\perp}$ and the signal amplitude is most impacted by out-of the plane spins $\chi_{1,2}^{\parallel}$. The effect of the spins in the gravitational waveform is commonly modelled  by the \textit{effective-spin precession parameter} $\chi_p$ and the \textit{effective-spin parameter} $\chi_{eff}$. These are expressed in terms of the component mass ratio $q=m_1/m_2\geq 1$~\cite{Will1995,Santamaria:2010yb, Schmidt:2014iyl} as
\begin{equation*}
\begin{aligned}
\chi_p=\max \left(\chi_{1}^\perp,\frac{2+3q/2}{q^2} \chi_{2}^\perp\right) ,\quad    \chi_{eff}=\frac{\chi_1^{||} m_1 + \chi_2^{||} m_2 }{m_1 + m_2}\,.
\end{aligned}
\label{spin}
\end{equation*}

Fig. \ref{fig:modes} shows that for an asymmetric and nearly edge-on system, an increase in $\chi_{eff}$ leads to a longer duration signal and hence louder signal. Also, a non-zero $\chi_p$ leads to a significant contribution from higher modes that are resulting in highly non-trivial waveform morphology. Since the PyCBC search implements template that model the dominant $(2,2)$ mode of aligned-spin sources, we expect it to be inefficient at detecting precessing/asymmetric sources as compared to the unmodelled cWB search.

\section{Search Algorithm}
\label{Search Algorithm}
In this section, we describe the two search algorithms used in NuRIA. The first one, PyCBC, is a matched-filter search that compares the incoming data to waveform templates for the quadrupole mode of aligned-spin BBHs. The second is a model-agnostic search for generic signals coherent across different detectors. Both the algorithms compute the significance of their signal candidates by ranking them together with the accidental background triggers according to a given ranking statistic. To estimate this background, the data of one detector is time-shifted by an amount which falls outside the physically viable time difference of the astrophysically coincident signal~\cite{GW150914-DETECTION}. This process, known as time-sliding, is repeated until a sufficient amount of background statistic is generated. The ranking statistic depends on the actual search and is tailored to provide a clear separation between background triggers and the signals targeted by the search. The final product of these searches is a list of trigger candidates with an associated astrophysical significance which is given by their inverse false alarm rate (IFAR). Triggers above a given IFAR threshold are then recorded as detections.

\subsection{Coherent WaveBurst}
\label{cWB}
Coherent WaveBurst~\cite{Klimenko:2015ypf} is an unmodelled, multi-detector, all-sky GW transient
search based on wavelet transform which looks for excess power in the
time-frequency domain. It targets a broad range of generic transient signals, with a minimal assumption about the underlying GW signal. An event is identified by clustering the time-frequency pixels with excess power as compared to the background noise level. Using the constrained maximum likelihood analysis method, the network correlation measures the correlation of the signal between the detectors, and the detection statistics ($\eta_c$) measures the signal-to-noise ratio. The events are then ranked based on network correlation and $\eta_c$, which help to distinguish the real GW from the non-Gaussian noise transients. It uses a large number of noise vetos to distinguish GW transients from noisy transients (for more details refer to Appendix A in~\cite{Gayathri:2019omo}). The noise-based vetoes rely on the residual noise energy per time-frequency pixel per detector and the extent of localisation of the noisy event in the time-frequency plane. Also, the signal based vetos are developed on the frequency evolution of the signal and the number of wavelets used for a given class of signal reconstruction~\cite{Tiwari:2015gal}. The veto values are tuned for IMBHB signals based on simulations study.
The cWB ranks candidate events that survived the cWB veto thresholds and are assigned a FAR value given by the rate of the corresponding background events with $\eta_c$ value more significant than that of the candidate event.

\subsection{The PyCBC search}

The PyCBC search matched-filters the incoming detector data $d$ with pre-computed waveform templates $h$. This filter is optimal when the template is a faithful representation of the GW signal present in the data. The output, known as the signal-to-noise ratio, is given by
\begin{equation}
    \rho^2 = 4 \left[\mathbb{R}e\int_{f_{low}}^{f_{high}} \frac{\tilde{d}^*(f)\tilde{h}(f)}{S_n(f)}df \right],
\end{equation}
where $\tilde{h}(f)$ denotes the Fourier transform of $h(t)$ (For details,~\cite{Owen:1998dk,1970esn..book.....W}). We set minimum and maximum frequencies $f_{low}=20$Hz and $f_{high}=2048$Hz, equal to the Nyquist frequency of our data. Coincident triggers across detectors with $\rho > 5.5$ are listed as signal candidate events and signal-template consistency vetoes are applied to these triggers to discriminate real GW signals from noisy transients of terrestrial origin known as \textit{glitches}~\cite{Canton:2013joa,Nitz:2017lco,Messick:2016aqy}. In particular, the PyCBC search implements a $\chi^2$ signal/glitch discriminator given by \cite{Allen:2005fk}
\begin{equation}
\chi^2_{r} = \frac{1}{2N-1}\sum_{i=1}^{N}(\rho_i-\rho_e)^2.
\end{equation}
Here, $\rho_i$ denotes the SNRs obtained in the i-th frequency band of the detector, chosen so that all of them are expected to produce equal SNR $\rho_e$ if the trigger perfectly matches the template $s$. If this is the case, then the veto statistic is expected to follow a $\chi^2$ distribution. Therefore, values close to unity are indicative of good signal/template consistency while a mismatch between them will lead to lower or larger values. Finally, the triggers are assigned a ranking statistic \cite{Davies:2020tsx, Usman:2015kfa,PhysRevD.87.024033,O1_O2_CATALOG}
\begin{equation}
\hat{\rho} =
\begin{cases}
 \frac{\rho}{[(1+(\chi^2_r)^{n/2})/2]^{1/n}} & \mathrm{for\,} \chi^2_r > 1 \\
 \rho & \mathrm{for\,} \chi^2_r \leq 1
\end{cases},
\label{eq:newsnr}
\end{equation}
which combines $\rho$ and $\chi^2_r$ \footnote{Normally $n=6$}. In addition, we also apply an additional test $\chi_{r,sg}^2$ to deal with instrumental glitches. It determines if the detector output contains any power greater than the maximum expected frequency content of a gravitational wave signal. Using this statistcs, we apply a further re-weighting as described in ~\cite{Nitz:2017lco} to obtain $\hat{\rho}_{new}$ to obtain single detector triggers. PyCBC then performs a coincidence test on the remaining triggers to obtain candidate events. These candidate events are assigned a ranking statistic which assesses their statistical significance and approximates the likelihood of obtaining the trigger parameters in the presence of a GW signal versus in the case of only noise~\cite{Nitz:2017svb}.  The significance of each of this ``foreground'' trigger is then estimated by comparing its $\hat{\rho}_{new}$ to the background distribution and is usually expressed in terms of inverse FAR in years (yr). For this study, we consider the same configuration of PyCBC used for the LIGO-Virgo O1-O2 IMBHB. The template bank~\cite{DalCanton:2017ala} targets the $(2,\pm2)$ modes of BBHs with total masses from $2 M_\odot$  to $500 M_\odot$, mass ratios up to $98$ and  restricted to spins aligned (anti-aligned) with the total angular momentum with maximum dimensionless aligned-spin parameter of 0.998. Additionally, the PyCBC algorithm excludes from the template banks templates shorter than $0.15s$ which are more easily mimicked by short glitches \cite{Canton2013}. This duration is defined as the time spanned by the template between the minimum frequency of the analysis ($20$Hz) and its ringdown frequency \cite{DalCanton:2017ala}. As pointed in \cite{DalCanton:2017ala}, we argue in Appendix I, this crucially affects the sensitivity of PyCBC at large masses. Finally, templates for BBHs heavier than $4M_\odot$ are computed with the reduced-order effective-one-body model SEOBNRv4ROM~\cite{Bohe:2016gbl}.

\section{Simulation Setup}
\label{Search Methodology}
\subsection{Injection Set}
We inject in the Advanced LIGO O1 data state-of-the-art numerically simulated signals for a large family of IMBHBs with generic spin configurations. Ideally, we would inject an astrophysically motivated population. However, this poses two main caveats. First, the true population of IMBHBs in
the Universe is unknown, preventing us from selecting a particular distribution. Second, numerical simulations
only cover a discrete set of spins and mass ratios space which prevents the study of a continuous set of sources.
We select 27 simulations with given mass ratios and spins, which can be then be scaled to arbitrary masses. These simulations are chosen so that they cover a reasonably wide range of mass ratios, and spin magnitudes and orientations parametrised by the $\chi_{eff}$ and $\chi_{p}$ parameters. We scale them to total masses in the range $M = 210 M_\odot, 300 M_\odot$ and $500 M_\odot$. Our particular choices are described in Table \ref{injections}. The numerical simulations have been computed by the Georgia Tech group (See Table \ref{injections} for a detailed list) using the {\tt Einstein Toolkit} code~\cite{2016CQGra..33t4001J, Zilhao:2013hia}) and are publicly available as part of the Georgia Tech Catalogue.  The waveforms include the modes
$= \{(2,\pm 1), (2,\pm 2), (3,\pm 2), (3 \pm 3),(4, \pm 2),(4, \pm 3), (4,\pm 4)\}$. We do not include further modes as they are usually very weak and dominated by numerical noise.
\begin{table}[bth]
\centering
    \begin{tabular}{|l|l|l|l|}
  \hline
      $\chi_p$ & $\chi_{eff}$   & q & SIM ID\\ \hline
     \multirow{8}{*}{0} & -0.6  & 2, 3 & GT0837,GT0846 \\
      & -0.4  & 1, 1.5, 2 & GT0564,GT0836,GT0838\\
      &   & 3 & GT0841 \\
    \cline{2-4}
       & 0  & 1,1.5,2 & GT0905,GT0477,GT0446 \\
      & 0  & 3,4,7 & GT0453,GT0454, GT0818\\\cline{2-4}
       & 0.4  & 1,1.5,2 & GT0422,GT0558,GT0472 \\
      & 0.3 & 3 & GT0596\\
      & 0.4, 0.45  & 2, 3 & GT0588,GT0600\\\hline
     \multirow{2}{*}{0.3} & -0.35  & 2 & GT0437 \\
     & 0.52  & 3 & GT0732 \\ \hline
     \multirow{3}{*}{0.42} & 0.51, 0.09 & 1,1.5 & GT0803,GT0873\\
     & 0.14, 0.21 & 2,3 & GT0872, GT0874 \\
      & 0.25, 0.32  & 4,7 & GT0875,GT0888 \\ \hline
     0.52 & -0.3  & 3 & GT0729 \\ \hline
     0.6 & 0 & 3 & GT0696
     \\ \hline
     \end{tabular}

    \caption{Summary of Georgia Tech NR simulations used to model our target signals. The source parameters are defined at the starting frequency of 16 Hz.}
    \label{injections}
\end{table}

For each of these, we create injection sets uniformly distributed over the sky sphere, uniformly distributed in the BBH orientation parameters ($\Phi, \cos{\iota}$), and uniformly distributed in co-moving volume up to a redshift of $z \approx 1$.

\subsection{Sensitive Distance Reach}

We determine the sensitivity of a search to each of our sources by calculating the corresponding sensitive distance reach. To do that, we inject a set of $N_{tot}$ injections distributed uniformly over the comoving volume $VT_{tot} [Gpc^3yr]$ into O1 data and recover them using our search algorithms. We consider as detections those recovered with significance equal or larger than a predetermined threshold~\cite{Abbott:2016drs, Abbott:2016nhf}. Denoting by $N_{rec}$ the number of detected injections, the corresponding sensitive volume and reach are computed as
\begin{equation}
    \label{volume-time sensitive}
    \braket{VT}_{sen}= \frac{N_{rec}}{N_{tot}} \braket{VT}_{tot} \,,
\end{equation}

\begin{equation}
    \label{sensitive distance reach}
    D_{\braket{VT}_{sen}}=\Big[\frac{3\braket{VT}_{sen}}{4 \pi T_a}\Big]^{1/3} \,,
\end{equation}
where $T_a$ is the total analysis time~\cite{O1O2IMBHBPaper}.
To compare the sensitivity of the two pipelines to the IMBH sources, we  compute the fractional difference in sensitive distance reach in percentage as
\begin{equation}
\label{relative improvement}
    \Delta D_{\braket{VT}_{sen}}[\%] = \left( \frac{D_{\braket{VT}_{sen}}^{cWB} -D_{\braket{VT}_{sen}}^{PyCBC}}{D_{\braket{VT}_{sen}}^{cWB}} \right)\times 100 .
\end{equation}
\linebreak
The positive values indicate better performance of cWB compared to PyCBC and vice-a-versa. We note that the statistical uncertainty in $N_{rec}$ is $\sigma({N_{rec})}=1/\sqrt{N_{rec}}$. Since cWB will be reasonably sensitive to all of our injection sets, we find that the corresponding sensitive distance uncertainties range in  $4-7\%$. However, for PyCBC the number of recovered injections can be as low as zero for the largest masses and mass ratios. For these cases, we will consider that a suitable statistical noise fluctuation would yield, at most, 1 recovered injection, giving an uncertaintiy of $100\%$.

Finally, for a given search, we place an astrophysical upper limit on the merger rate density at the 90 \% confidence level as obtained by \cite{Biswas:2007ni, O1O2IMBHBPaper}:
\begin{equation} \label{rate}
    R_{90 \%} = -\frac{\ln{0.1} }{\braket{VT}_{sen}} \,.
\end{equation}

\section{Results} \label{Results}
In this section, we first compare the sensitivity of the two search algorithms using a fraction of O1 data (between September 12 - October 8, 2015) and find that cWB largely over-performs PyCBC in most cases. Next, we place upper limits on the coalescence rate density for a precessing set using O1 data using the results of the cWB search.

\subsection{Comparing the searches}
We compare the sensitivity of our searches at two reference significance thresholds given by IFARs of 2.94yr
\footnote{This is motivated by loudest IMBH-like trigger reported in~\cite{O1O2IMBHBPaper}} and 300yr. At low IFAR, the significance of the PyCBC triggers is mostly given by the recovered SNR so that a good separation of injections from the background is not required. Hence, subtle physical effects that cause a signal-template mismatch may not play a role in the search comparison.

At a larger IFAR of 300yr, the worsening effect comes because of mismatch between the signals and the quadrupolar templates and hence large deviations of $\chi^2$ from unity.
Fig.~\ref{fig:Comparison between the two pipelines} shows $\Delta D _{\braket{VT}_{sen}}$, at an IFAR threshold of 2.94yr, for all the sources considered in this study, expressed in the $(m_1,m_2)$ plane, with varying $\chi_p$ and $\chi_{eff}$. For most cases, cWB out-performs PyCBC, so that $\Delta D _{\braket{VT}_{sen}}>0$. In agreement with previous studies restricted to aligned-spins~\cite{O1O2IMBHBPaper}, the difference between the two searches increases with an increase in total mass for fixed mass-ratio and spin parameters. This is partially due to the increasing contribution of higher-modes to the signals, not included in the PyCBC search templates. On the one hand, the mismatch between injections and templates leads to a poor SNR recovery. (For details check ~\citep{CalderonBustillo:2017skv}). On the other, it increases the $\chi^2_{r}$ statistic, making the search interpret the injections as glitches.  Additionally, we note that even in the absence of higher-modes, it has been shown in the past that the $\chi_{r}^2$ discriminator performs poorly at separating short-duration signals from glitches \cite{Nitz:2017lco, Dhurandhar:2017aan}.

We note that, in a somewhat unexpected way, for bin $(M = 500 M_\odot, q = 2)$ PyCBC was not able to recover any injection while some injections are recovered for the neighbouring $q=1.5$ and $q=3$ cases. We attribute this to a combination of the large statistical uncertainty of the PyCBC sensitive range and the physical properties of the sources. First, the very low number of recovered injections $N_{rec} \in[1,5] $ at such large masses yields relatively large uncertainties $\sigma(N_{rec})=1/\sqrt{N_{rec}}$. Second, at such large masses, the $(2,2)$ mode of the system can be mostly out of the sensitive band so that the PyCBC templates can effectively match the next mode remaining in the band, namely the $(3,3)$ instead of having to match a combination of modes. Such modes are more prominent for increasing mass ratio. Finally, note that performance of PyCBC relative to cWB improves for larger mass ratios due the decreasing sensitivity of cWB, shown in Fig.4(a). A similar effect is also noticeable in Fig.~\ref{fig:Comparison between the two pipelines}(c). A detailed description about the physical reason for missing the mass-bin is given in Appendix~\ref{Appendix A}.

At a larger IFAR, mismatches between injections and templates affect the sensitivity of PyCBC. As a consequence, there is a reduction in its sensitivity toward high total mass and high mass ratio sources with more considerable higher mode contribution. Consistently, even at this IFAR (see Fig.  (\ref{fig:Comparison between the two pipelines at IFAR 300})) we find that the two pipelines have a comparable performance for low mass and low mass ratio systems, as the impact of precession / higher modes on the signals being less important in these cases.

We conclude that, as expected, the signal morphology of IMBHB sources -- \textit{short signals with potentially complex morphology in the case of edge-on, large mass ratio cases} -- is better captured by the model agnostic cWB search than the by the PyCBC search. In the following, we report the simulation results with cWB using the entire duration of O1.

\subsection{IMBHBs with generic spins}
We estimate the sensitive distance reach using Eq.\ref{sensitive distance reach}.
As a general trend, we observe that the sensitivity decreases with increasing total mass and mass-ratio. This is expected as the duration of signal shortens in the detector band, the loudness of the signal decreases. As mentioned before, a positive $\chi_{eff}$ leads to a longer signal. For this reason, in Fig.~\ref{fig:cWB} (b), we observe a larger sensitive volume for $q=3$, than for the $q=2$ case, which has negative $\chi_{eff}$.  For this same reason, the cases with $(\chi_p,\chi_{eff}) = (0.4)$ and positive $\chi_{eff}$ in Fig.~\ref{fig:Comparison between the two pipelines at IFAR 300}(c) shows the largest distance reach among all sources.

We compute the corresponding $90 \%$ merger rate density using eq. \eqref{rate}. These are shown in Fig. \ref{fig:Rate_cWB}. We place the most constraining upper limit of $0.36/Gpc^3/yr$ on the merger rate of equal-mass IMBHs with $M=210M_\odot$ and effective spin parameters $(\chi_p,\chi_{eff}) = (0.4,0.51)$. We note this upper limit improves by a factor of $\sim 3$ on the one obtained for aligned-spin sources after the first Advanced LIGO observing run \cite{O1-IMBHB}. This is less constraining than the one obtained after the second run before which the detectors underwent major up-gradation to obtain better sensitivity. Better sensitivity shows that generically spinning sources may offer a better chance to observe BBHs in this mass range.

\subsection{Precession vs. aligned spins}
It is natural to ask if the effects of precession and aligned-spins, parametrised respectively by $\chi_p$ and $\chi_{eff}$, can be somewhat disentangled in terms of the sensitivity. To do this, we compute the sensitivity to sources with fixed mass-ratio and total mass; with $\chi_p=0$ and varying $\chi_{eff}$, and varying $\chi_p$ for fixed $\chi_{eff}$. The right panels of Fig. \ref{fig:generic spin} show the results for the case of $q=3$. As expected, for fixed $\chi_p = 0$, a positive (negative) $\chi_{eff}$ leads to larger (lower) sensitive distance reach. Besides, we observe that a variation of  $\chi_p$ produces a variation between $4\%$ to $7\%$ in the sensitive range for a fixed $\chi_{eff}$. Similar results are observed for the $q=2$ cases shown in the left panels. Given this, we conclude that for these mass ratios, the sensitive distance reach is not significantly affected by the value of $\chi_p$.

\subsection{Impact of higher-order modes}
Finally, similar to what was done in \cite{CalderonBustillo:2017skv} for aligned-spin sources, we look at the impact of the inclusion/omission of the higher modes. To this, we compare the sensitivity of our search to injection sets including and omitting this effect. Fig.\ref{fig:(2,2) mode vs HM} shows the fractional increase of sensitive distance produced by the inclusion of higher-order modes in our injections. We observe that the sensitivity of the pipeline increases when the higher-modes are included in the injections, as this generally increases the available signal power. Since higher-modes have a larger impact on the case of large mass-ratio and large total-mass sources, the impact in the sensitive distance reach is larger in these cases. An increment as large as  $\sim{57 \%}$ is observed for the system with $\chi_{p}=0.42$, $M=500 M_\odot$ and $q=7$. An apparent deviation from this rule can be observed in panel b, in which the impact of HMs is larger for the $q=2$ than for the $q=3$ case. The reason is that while the $q=3$ case has a $\chi_{eff} >0$ that makes the signal long in the detector sensitive band, the $q=2$ case has $\chi_{eff} < 0$. The latter leads to a short observable signal more dominated by its merger and ringdown portions, in which HMs are more prominent. Finally, we note that negative values apparently indicating that a larger sensitivity is obtained when HMs are omitted, are statistically consistent with zero.

\section{Conclusion}
\label{Conclusion}
The detection of intermediate-mass black holes is a standing challenge in astronomy. Despite being one of the loudest sources for advanced gravitational-wave detectors, the shorter duration of the signals in the detector sensitive band and the prominent impact of higher modes and possibly precession (not captured by model-based searches) makes their detection more difficult than that of lighter binary black holes. In this situation, un-modelled searches have shown to be a promising method toward the detection of such objects~\cite{O1O2IMBHBPaper}. In NuRIA, for the first time, we present a comprehensive study on the ability of current gravitational-wave searches to detect generic spinning IMBHBs. We focus on two searches used by the LIGO-Virgo collaborations in their recent second observing runs: the matched-filter algorithm PyCBC and model agnostic cWB. We find that at their present status, the latter offers a much better prospect to observe IMBHBs in GW window. Finally, we have placed the first-ever upper limits on the coalescence rate of precessing IMBHs using data from the first Advanced LIGO observing run using the un-modelled search.
We place our most stringent $90 \%$ coalescence rate density upper limit of $0.36/Gpc^3/yr$ for equal mass and precessing system with a total mass of $210 M_\odot$. This improves on the $0.94/Gpc^3/yr$ placed for aligned-spin IMBHs after the first Advanced LIGO observing run. The higher sensitive distance reach of precessing systems indicate that generically spinning sources offer a better chance for the detection of BBHs in this mass range. While the upper limit on rate density has been pushed to $0.2/Gpc^3/yr$ after the second observing run, we expect that more constraining limits when these are computed using injections from generically spinning binaries as well as when the detectors become more and more sensitive.

\section{Acknowledgements}
The authors thank Grace Kim for sharing with us a detailed comparison between Georgia Tech and SXS waveforms. The authors are grateful for the computational resources provided by the LIGO Laboratory and supported by National Science Foundation Grants PHY-0757058 and PHY-0823459. This research has made use of data obtained from the Gravitational Wave Open Science Center (https://www.gwopenscience.org)~\cite{OpenData}, a service of LIGO Laboratory, the LIGO Scientific Collaboration and the Virgo Collaboration. We also thank the LIGO-VIRGO IMBHB group, the PyCBC and the cWB team for their help and support, especially Sebastian Khan, Ian Harry, Sergey Klimenko, Imre Bartos, Leslie Wade and Thomas Dent. KC acknowledges the MHRD, Government of India for the fellowship support. VG recognises Inspire division, Department of Science and Technology, Government of India for the fellowship support. J.C.B  is supported by the Australian Research Council Discovery Project DP180103155 and by the Direct Grant, Project 4053406, from the Research Committee of the Chinese University of Hong Kong. AP thanks the IRCC, SEED grant, IIT Bombay for the support. This document has LIGO DCC number P1900366.

\begin{figure}[h!]
    \centering
    \includegraphics[width=0.47\textwidth]{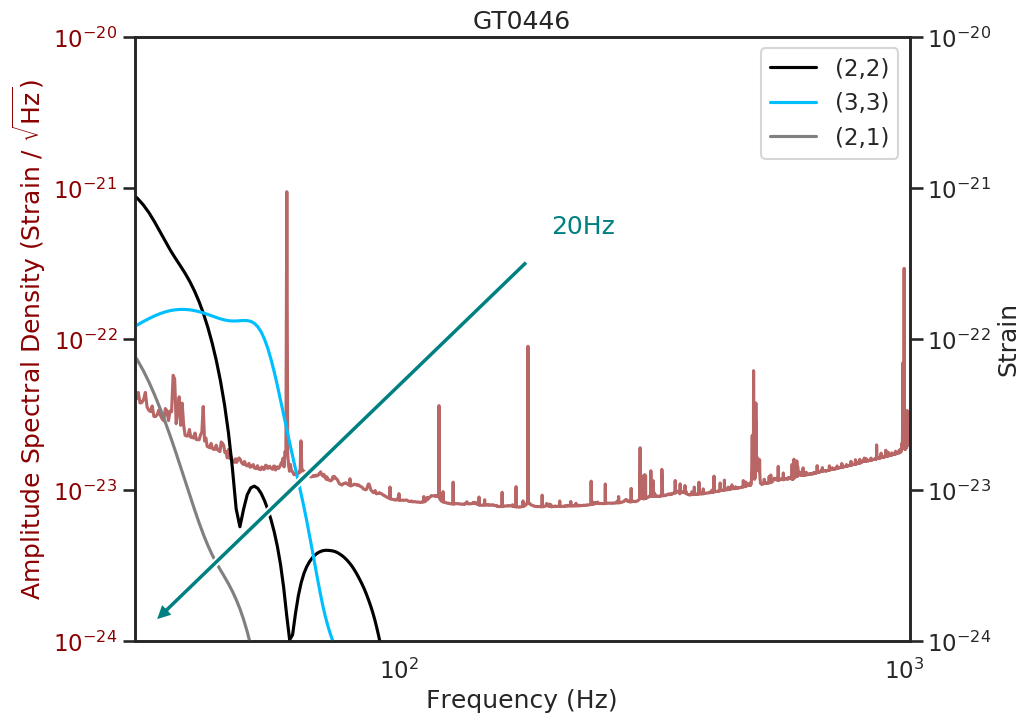}
     \includegraphics[width=0.47\textwidth]{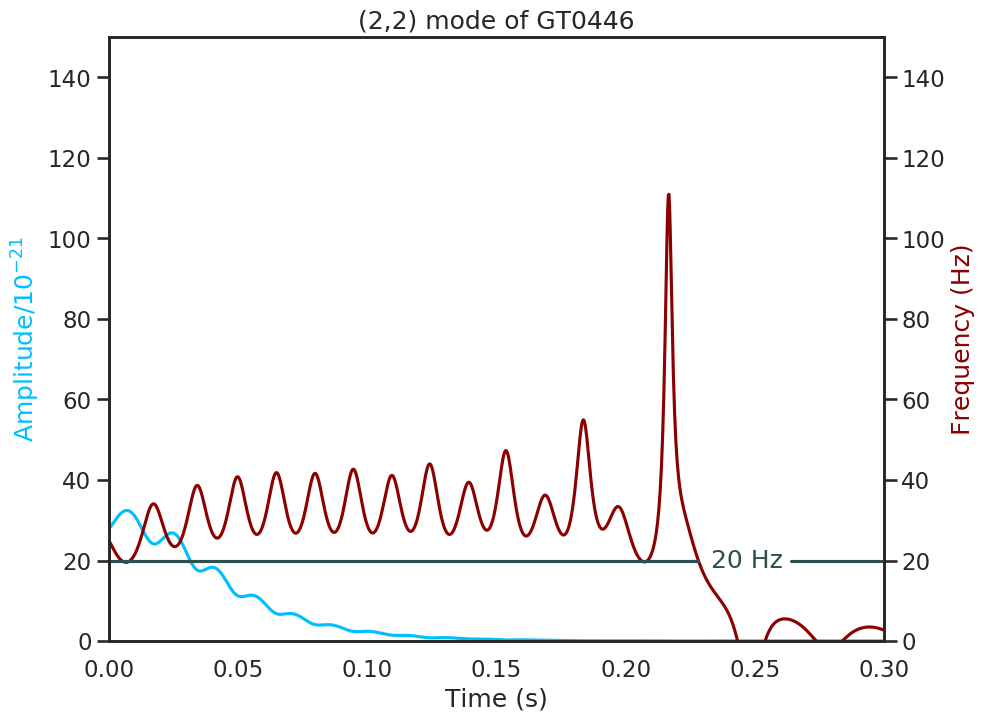}
    \caption{Top: Amplitude of the Fourier transforms of the $(2,2)$ and $(3,3)$ modes of the GT0446 numerical simulation scaled to total mass of $500 M_\odot$ at a distance of 100Mpc. We have chosen an inclination $\iota=60^\circ$ to make the $(3,3)$ mode sufficiently prominent. Bottom: Frequency and amplitude of the $(2,2)$ as a function of time.}
    \label{fig:GT0446}
\end{figure}
\appendix
\section{Investigating PyCBC's poor performance at high masses}\label{Appendix A}
In this appendix we provide visual intuition on the poor performance of PyCBC for large masses. The top panel of Fig.~\ref{fig:GT0446}, shows the spectrum of the $(2,2)$ and $(3,3)$ modes of the non-spinning $q=2$ simulation GT0446, scaled to $500M_\odot$, for which PyCBC could not recover any injection. The bottom panel, shows the frequency and amplitude of the dominant $(2,2)$ as a function of time. We note that oscillations in the frequency become more pronounce as the signal becomes weaker and the simulation is dominated by numerical noise.\\

First, the top panel shows that even though the $(3,3)$ mode spans a larger range of frequencies above the Advanced LIGO noise, the $(2,2)$ mode is an order of magnitude stronger, so the presence of higher modes does not seem to be the cause of PyCBC's poor performance. However, the bottom panel shows that the time spanned by the $(2,2)$ after reaching $20$Hz is below the minimum threshold of $0.15$s template duration imposed by the PyCBC template bank ~\cite{DalCanton:2017ala}. Moreover, we note that while this threshold refers to the time spanned by the template between $20$Hz and its ringdown frequency, the duration of the \textit{full waveform} within the analysis frequency band is actually shorter than $0.15$s. Consequently, even if a putative template including only the $(2,2)$-mode could recover the signal, this would be automatically discarded by PyCBC.  While PyCBC uses this strategy to avoid the accumulation of large amount of triggers caused by short glitches \cite{Canton2013}, this decreases the ability of the template-bank based matched-filter search to recover IMBHB signals.



\bibliography{reference}

\begin{thebibliography}{70}%
\makeatletter
\providecommand \@ifxundefined [1]{%
 \@ifx{#1\undefined}
}%
\providecommand \@ifnum [1]{%
 \ifnum #1\expandafter \@firstoftwo
 \else \expandafter \@secondoftwo
 \fi
}%
\providecommand \@ifx [1]{%
 \ifx #1\expandafter \@firstoftwo
 \else \expandafter \@secondoftwo
 \fi
}%
\providecommand \natexlab [1]{#1}%
\providecommand \enquote  [1]{``#1''}%
\providecommand \bibnamefont  [1]{#1}%
\providecommand \bibfnamefont [1]{#1}%
\providecommand \citenamefont [1]{#1}%
\providecommand \href@noop [0]{\@secondoftwo}%
\providecommand \href [0]{\begingroup \@sanitize@url \@href}%
\providecommand \@href[1]{\@@startlink{#1}\@@href}%
\providecommand \@@href[1]{\endgroup#1\@@endlink}%
\providecommand \@sanitize@url [0]{\catcode `\\12\catcode `\$12\catcode
  `\&12\catcode `\#12\catcode `\^12\catcode `\_12\catcode `\%12\relax}%
\providecommand \@@startlink[1]{}%
\providecommand \@@endlink[0]{}%
\providecommand \url  [0]{\begingroup\@sanitize@url \@url }%
\providecommand \@url [1]{\endgroup\@href {#1}{\urlprefix }}%
\providecommand \urlprefix  [0]{URL }%
\providecommand \Eprint [0]{\href }%
\providecommand \doibase [0]{http://dx.doi.org/}%
\providecommand \selectlanguage [0]{\@gobble}%
\providecommand \bibinfo  [0]{\@secondoftwo}%
\providecommand \bibfield  [0]{\@secondoftwo}%
\providecommand \translation [1]{[#1]}%
\providecommand \BibitemOpen [0]{}%
\providecommand \bibitemStop [0]{}%
\providecommand \bibitemNoStop [0]{.\EOS\space}%
\providecommand \EOS [0]{\spacefactor3000\relax}%
\providecommand \BibitemShut  [1]{\csname bibitem#1\endcsname}%
\let\auto@bib@innerbib\@empty
\bibitem [{\citenamefont {Aasi}\ \emph {et~al.}(2015)\citenamefont {Aasi} \emph
  {et~al.}}]{aLIGODetector}%
  \BibitemOpen
  \bibfield  {author} {\bibinfo {author} {\bibfnamefont {J.}~\bibnamefont
  {Aasi}} \emph {et~al.} (\bibinfo {collaboration} {LIGO Scientific}),\ }\href
  {\doibase 10.1088/0264-9381/32/7/074001} {\bibfield  {journal} {\bibinfo
  {journal} {Class. Quant. Grav.}\ }\textbf {\bibinfo {volume} {32}},\ \bibinfo
  {pages} {074001} (\bibinfo {year} {2015})},\ \Eprint
  {http://arxiv.org/abs/1411.4547} {arXiv:1411.4547 [gr-qc]} \BibitemShut
  {NoStop}%
\bibitem [{\citenamefont {Acernese}\ \emph {et~al.}(2015)\citenamefont
  {Acernese} \emph {et~al.}}]{TheVirgo:2014hva}%
  \BibitemOpen
  \bibfield  {author} {\bibinfo {author} {\bibfnamefont {F.}~\bibnamefont
  {Acernese}} \emph {et~al.} (\bibinfo {collaboration} {VIRGO}),\ }\href
  {\doibase 10.1088/0264-9381/32/2/024001} {\bibfield  {journal} {\bibinfo
  {journal} {Class. Quant. Grav.}\ }\textbf {\bibinfo {volume} {32}},\ \bibinfo
  {pages} {024001} (\bibinfo {year} {2015})},\ \Eprint
  {http://arxiv.org/abs/1408.3978} {arXiv:1408.3978 [gr-qc]} \BibitemShut
  {NoStop}%
\bibitem [{\citenamefont {Acernese}\ \emph {et~al.}(2018)\citenamefont
  {Acernese} \emph {et~al.}}]{Acernese_2018}%
  \BibitemOpen
  \bibfield  {author} {\bibinfo {author} {\bibfnamefont {F.}~\bibnamefont
  {Acernese}} \emph {et~al.} (\bibinfo {collaboration} {Virgo}),\ }\href
  {\doibase 10.1088/1361-6382/aadf1a} {\bibfield  {journal} {\bibinfo
  {journal} {Class. Quant. Grav.}\ }\textbf {\bibinfo {volume} {35}},\ \bibinfo
  {pages} {205004} (\bibinfo {year} {2018})},\ \Eprint
  {http://arxiv.org/abs/1807.03275} {arXiv:1807.03275 [gr-qc]} \BibitemShut
  {NoStop}%
\bibitem [{\citenamefont {Abbott}\ \emph
  {et~al.}(2017{\natexlab{a}})\citenamefont {Abbott} \emph
  {et~al.}}]{GW170817-DETECTION}%
  \BibitemOpen
  \bibfield  {author} {\bibinfo {author} {\bibfnamefont {B.}~\bibnamefont
  {Abbott}} \emph {et~al.} (\bibinfo {collaboration} {LIGO Scientific
  Collaboration, Virgo Collaboration}),\ }\href {\doibase
  10.1103/PhysRevLett.119.161101} {\bibfield  {journal} {\bibinfo  {journal}
  {Phys. Rev. Lett.}\ }\textbf {\bibinfo {volume} {119}},\ \bibinfo {pages}
  {161101} (\bibinfo {year} {2017}{\natexlab{a}})},\ \Eprint
  {http://arxiv.org/abs/1710.05832} {arXiv:1710.05832 [gr-qc]} \BibitemShut
  {NoStop}%
\bibitem [{\citenamefont {Abbott}\ \emph
  {et~al.}(2016{\natexlab{a}})\citenamefont {Abbott} \emph
  {et~al.}}]{GW150914-DETECTION}%
  \BibitemOpen
  \bibfield  {author} {\bibinfo {author} {\bibfnamefont {B.~P.}\ \bibnamefont
  {Abbott}} \emph {et~al.} (\bibinfo {collaboration} {LIGO Scientific
  Collaboration, Virgo Collaboration}),\ }\href {\doibase
  10.1103/PhysRevLett.116.061102} {\bibfield  {journal} {\bibinfo  {journal}
  {Phys. Rev. Lett.}\ }\textbf {\bibinfo {volume} {116}},\ \bibinfo {pages}
  {061102} (\bibinfo {year} {2016}{\natexlab{a}})},\ \Eprint
  {http://arxiv.org/abs/1602.03837} {arXiv:1602.03837 [gr-qc]} \BibitemShut
  {NoStop}%
\bibitem [{\citenamefont {Abbott}\ \emph
  {et~al.}(2016{\natexlab{b}})\citenamefont {Abbott} \emph
  {et~al.}}]{GW151226-DETECTION}%
  \BibitemOpen
  \bibfield  {author} {\bibinfo {author} {\bibfnamefont {B.~P.}\ \bibnamefont
  {Abbott}} \emph {et~al.} (\bibinfo {collaboration} {LIGO Scientific
  Collaboration, Virgo Collaboration}),\ }\href {\doibase
  10.1103/PhysRevLett.116.241103} {\bibfield  {journal} {\bibinfo  {journal}
  {Phys. Rev. Lett.}\ }\textbf {\bibinfo {volume} {116}},\ \bibinfo {pages}
  {241103} (\bibinfo {year} {2016}{\natexlab{b}})},\ \Eprint
  {http://arxiv.org/abs/1606.04855} {arXiv:1606.04855 [gr-qc]} \BibitemShut
  {NoStop}%
\bibitem [{\citenamefont {Abbott}\ \emph
  {et~al.}(2017{\natexlab{b}})\citenamefont {Abbott} \emph
  {et~al.}}]{GW170104-DETECTION}%
  \BibitemOpen
  \bibfield  {author} {\bibinfo {author} {\bibfnamefont {B.~P.}\ \bibnamefont
  {Abbott}} \emph {et~al.} (\bibinfo {collaboration} {LIGO Scientific
  Collaboration, Virgo Collaboration}),\ }\href {\doibase
  10.1103/PhysRevLett.118.221101} {\bibfield  {journal} {\bibinfo  {journal}
  {Phys. Rev. Lett.}\ }\textbf {\bibinfo {volume} {118}},\ \bibinfo {pages}
  {221101} (\bibinfo {year} {2017}{\natexlab{b}})},\ \Eprint
  {http://arxiv.org/abs/1706.01812} {arXiv:1706.01812 [gr-qc]} \BibitemShut
  {NoStop}%
\bibitem [{\citenamefont {Abbott}\ \emph
  {et~al.}(2017{\natexlab{c}})\citenamefont {Abbott} \emph
  {et~al.}}]{GW170608-DETECTION}%
  \BibitemOpen
  \bibfield  {author} {\bibinfo {author} {\bibfnamefont {B.~P.}\ \bibnamefont
  {Abbott}} \emph {et~al.} (\bibinfo {collaboration} {LIGO Scientific
  Collaboration, Virgo Collaboration}),\ }\href {\doibase
  10.3847/2041-8213/aa9f0c} {\bibfield  {journal} {\bibinfo  {journal}
  {Astrophys. J.}\ }\textbf {\bibinfo {volume} {851}},\ \bibinfo {pages} {L35}
  (\bibinfo {year} {2017}{\natexlab{c}})},\ \Eprint
  {http://arxiv.org/abs/1711.05578} {arXiv:1711.05578 [astro-ph.HE]}
  \BibitemShut {NoStop}%
\bibitem [{\citenamefont {Abbott}\ \emph
  {et~al.}(2017{\natexlab{d}})\citenamefont {Abbott} \emph
  {et~al.}}]{GW170814-DETECTION}%
  \BibitemOpen
  \bibfield  {author} {\bibinfo {author} {\bibfnamefont {B.~P.}\ \bibnamefont
  {Abbott}} \emph {et~al.} (\bibinfo {collaboration} {LIGO Scientific
  Collaboration, Virgo Collaboration}),\ }\href {\doibase
  10.1103/PhysRevLett.119.141101} {\bibfield  {journal} {\bibinfo  {journal}
  {Phys. Rev. Lett.}\ }\textbf {\bibinfo {volume} {119}},\ \bibinfo {pages}
  {141101} (\bibinfo {year} {2017}{\natexlab{d}})},\ \Eprint
  {http://arxiv.org/abs/1709.09660} {arXiv:1709.09660 [gr-qc]} \BibitemShut
  {NoStop}%
\bibitem [{\citenamefont {Abbott}\ \emph
  {et~al.}(2016{\natexlab{c}})\citenamefont {Abbott} \emph
  {et~al.}}]{TheLIGOScientific:2016pea}%
  \BibitemOpen
  \bibfield  {author} {\bibinfo {author} {\bibfnamefont {B.~P.}\ \bibnamefont
  {Abbott}} \emph {et~al.} (\bibinfo {collaboration} {LIGO Scientific
  Collaboration, Virgo Collaboration}),\ }\href {\doibase
  10.1103/PhysRevX.6.041015, 10.1103/PhysRevX.8.039903} {\bibfield  {journal}
  {\bibinfo  {journal} {Phys. Rev.}\ }\textbf {\bibinfo {volume} {X6}},\
  \bibinfo {pages} {041015} (\bibinfo {year} {2016}{\natexlab{c}})},\ \bibinfo
  {note} {[erratum: Phys. Rev.X8,no.3,039903(2018)]},\ \Eprint
  {http://arxiv.org/abs/1606.04856} {arXiv:1606.04856 [gr-qc]} \BibitemShut
  {NoStop}%
\bibitem [{\citenamefont {Abbott}\ \emph
  {et~al.}(2019{\natexlab{a}})\citenamefont {Abbott} \emph
  {et~al.}}]{O1_O2_CATALOG}%
  \BibitemOpen
  \bibfield  {author} {\bibinfo {author} {\bibfnamefont {B.~P.}\ \bibnamefont
  {Abbott}} \emph {et~al.} (\bibinfo {collaboration} {LIGO Scientific,
  Virgo}),\ }\href {\doibase 10.1103/PhysRevX.9.031040} {\bibfield  {journal}
  {\bibinfo  {journal} {Phys. Rev.}\ }\textbf {\bibinfo {volume} {X9}},\
  \bibinfo {pages} {031040} (\bibinfo {year} {2019}{\natexlab{a}})},\ \Eprint
  {http://arxiv.org/abs/1811.12907} {arXiv:1811.12907 [astro-ph.HE]}
  \BibitemShut {NoStop}%
\bibitem [{\citenamefont {Nitz}\ \emph
  {et~al.}(2019{\natexlab{a}})\citenamefont {Nitz}, \citenamefont {Capano},
  \citenamefont {Nielsen}, \citenamefont {Reyes}, \citenamefont {White},
  \citenamefont {Brown},\ and\ \citenamefont {Krishnan}}]{1-OGC}%
  \BibitemOpen
  \bibfield  {author} {\bibinfo {author} {\bibfnamefont {A.~H.}\ \bibnamefont
  {Nitz}}, \bibinfo {author} {\bibfnamefont {C.}~\bibnamefont {Capano}},
  \bibinfo {author} {\bibfnamefont {A.~B.}\ \bibnamefont {Nielsen}}, \bibinfo
  {author} {\bibfnamefont {S.}~\bibnamefont {Reyes}}, \bibinfo {author}
  {\bibfnamefont {R.}~\bibnamefont {White}}, \bibinfo {author} {\bibfnamefont
  {D.~A.}\ \bibnamefont {Brown}}, \ and\ \bibinfo {author} {\bibfnamefont
  {B.}~\bibnamefont {Krishnan}},\ }\href {\doibase 10.3847/1538-4357/ab0108}
  {\bibfield  {journal} {\bibinfo  {journal} {Astrophys. J.}\ }\textbf
  {\bibinfo {volume} {872}},\ \bibinfo {pages} {195} (\bibinfo {year}
  {2019}{\natexlab{a}})},\ \Eprint {http://arxiv.org/abs/1811.01921}
  {arXiv:1811.01921 [gr-qc]} \BibitemShut {NoStop}%
\bibitem [{\citenamefont {Nitz}\ \emph
  {et~al.}(2019{\natexlab{b}})\citenamefont {Nitz}, \citenamefont {Dent},
  \citenamefont {Davies}, \citenamefont {Kumar}, \citenamefont {Capano},
  \citenamefont {Harry}, \citenamefont {Mozzon}, \citenamefont {Nuttall},
  \citenamefont {Lundgren},\ and\ \citenamefont {Tápai}}]{2-OGC}%
  \BibitemOpen
  \bibfield  {author} {\bibinfo {author} {\bibfnamefont {A.~H.}\ \bibnamefont
  {Nitz}}, \bibinfo {author} {\bibfnamefont {T.}~\bibnamefont {Dent}}, \bibinfo
  {author} {\bibfnamefont {G.~S.}\ \bibnamefont {Davies}}, \bibinfo {author}
  {\bibfnamefont {S.}~\bibnamefont {Kumar}}, \bibinfo {author} {\bibfnamefont
  {C.~D.}\ \bibnamefont {Capano}}, \bibinfo {author} {\bibfnamefont
  {I.}~\bibnamefont {Harry}}, \bibinfo {author} {\bibfnamefont
  {S.}~\bibnamefont {Mozzon}}, \bibinfo {author} {\bibfnamefont
  {L.}~\bibnamefont {Nuttall}}, \bibinfo {author} {\bibfnamefont
  {A.}~\bibnamefont {Lundgren}}, \ and\ \bibinfo {author} {\bibfnamefont
  {M.}~\bibnamefont {Tápai}},\ }\href {\doibase 10.3847/1538-4357/ab733f}
  {\bibfield  {journal} {\bibinfo  {journal} {Astrophys. J.}\ }\textbf
  {\bibinfo {volume} {891}},\ \bibinfo {pages} {123} (\bibinfo {year}
  {2019}{\natexlab{b}})},\ \Eprint {http://arxiv.org/abs/1910.05331}
  {arXiv:1910.05331 [astro-ph.HE]} \BibitemShut {NoStop}%
\bibitem [{\citenamefont {Venumadhav}\ \emph
  {et~al.}(2019{\natexlab{a}})\citenamefont {Venumadhav}, \citenamefont
  {Zackay}, \citenamefont {Roulet}, \citenamefont {Dai},\ and\ \citenamefont
  {Zaldarriaga}}]{IAS-Catalogue}%
  \BibitemOpen
  \bibfield  {author} {\bibinfo {author} {\bibfnamefont {T.}~\bibnamefont
  {Venumadhav}}, \bibinfo {author} {\bibfnamefont {B.}~\bibnamefont {Zackay}},
  \bibinfo {author} {\bibfnamefont {J.}~\bibnamefont {Roulet}}, \bibinfo
  {author} {\bibfnamefont {L.}~\bibnamefont {Dai}}, \ and\ \bibinfo {author}
  {\bibfnamefont {M.}~\bibnamefont {Zaldarriaga}},\ }\href {\doibase
  10.1103/PhysRevD.100.023011} {\bibfield  {journal} {\bibinfo  {journal}
  {Phys. Rev.}\ }\textbf {\bibinfo {volume} {D100}},\ \bibinfo {pages} {023011}
  (\bibinfo {year} {2019}{\natexlab{a}})},\ \Eprint
  {http://arxiv.org/abs/1902.10341} {arXiv:1902.10341 [astro-ph.IM]}
  \BibitemShut {NoStop}%
\bibitem [{\citenamefont {Venumadhav}\ \emph
  {et~al.}(2019{\natexlab{b}})\citenamefont {Venumadhav}, \citenamefont
  {Zackay}, \citenamefont {Roulet}, \citenamefont {Dai},\ and\ \citenamefont
  {Zaldarriaga}}]{IAS-Catalogue_II}%
  \BibitemOpen
  \bibfield  {author} {\bibinfo {author} {\bibfnamefont {T.}~\bibnamefont
  {Venumadhav}}, \bibinfo {author} {\bibfnamefont {B.}~\bibnamefont {Zackay}},
  \bibinfo {author} {\bibfnamefont {J.}~\bibnamefont {Roulet}}, \bibinfo
  {author} {\bibfnamefont {L.}~\bibnamefont {Dai}}, \ and\ \bibinfo {author}
  {\bibfnamefont {M.}~\bibnamefont {Zaldarriaga}},\ }\href@noop {} {\
  (\bibinfo {year} {2019}{\natexlab{b}})},\ \Eprint
  {http://arxiv.org/abs/1904.07214} {arXiv:1904.07214 [astro-ph.HE]}
  \BibitemShut {NoStop}%
\bibitem [{\citenamefont {Antelis}\ and\ \citenamefont
  {Moreno}(2019)}]{Antelis:2018smo}%
  \BibitemOpen
  \bibfield  {author} {\bibinfo {author} {\bibfnamefont {J.~M.}\ \bibnamefont
  {Antelis}}\ and\ \bibinfo {author} {\bibfnamefont {C.}~\bibnamefont
  {Moreno}},\ }\href {\doibase 10.1007/s10714-019-2546-x} {\bibfield  {journal}
  {\bibinfo  {journal} {Gen. Rel. Grav.}\ }\textbf {\bibinfo {volume} {51}},\
  \bibinfo {pages} {61} (\bibinfo {year} {2019})},\ \Eprint
  {http://arxiv.org/abs/1807.07660} {arXiv:1807.07660 [gr-qc]} \BibitemShut
  {NoStop}%
\bibitem [{\citenamefont {Abbott}\ \emph {et~al.}(2020)\citenamefont {Abbott}
  \emph {et~al.}}]{GW190425-Discovery}%
  \BibitemOpen
  \bibfield  {author} {\bibinfo {author} {\bibfnamefont {B.~P.}\ \bibnamefont
  {Abbott}} \emph {et~al.} (\bibinfo {collaboration} {LIGO Scientific,
  Virgo}),\ }\href {\doibase 10.3847/2041-8213/ab75f5} {\bibfield  {journal}
  {\bibinfo  {journal} {Astrophys. J. Lett.}\ }\textbf {\bibinfo {volume}
  {892}},\ \bibinfo {pages} {L3} (\bibinfo {year} {2020})},\ \Eprint
  {http://arxiv.org/abs/2001.01761} {arXiv:2001.01761 [astro-ph.HE]}
  \BibitemShut {NoStop}%
\bibitem [{\citenamefont {Eardley}\ and\ \citenamefont
  {Press}(1975)}]{EardleyPress1975}%
  \BibitemOpen
  \bibfield  {author} {\bibinfo {author} {\bibfnamefont {D.~M.}\ \bibnamefont
  {Eardley}}\ and\ \bibinfo {author} {\bibfnamefont {W.~H.}\ \bibnamefont
  {Press}},\ }\href {\doibase 10.1146/annurev.aa.13.090175.002121} {\bibfield
  {journal} {\bibinfo  {journal} {Annual Review of Astronomy and Astrophysics}\
  }\textbf {\bibinfo {volume} {13}},\ \bibinfo {pages} {381} (\bibinfo {year}
  {1975})},\ \Eprint
  {http://arxiv.org/abs/https://doi.org/10.1146/annurev.aa.13.090175.002121}
  {https://doi.org/10.1146/annurev.aa.13.090175.002121} \BibitemShut {NoStop}%
\bibitem [{\citenamefont {{Rees}}(1978)}]{Rees1978}%
  \BibitemOpen
  \bibfield  {author} {\bibinfo {author} {\bibfnamefont {M.~J.}\ \bibnamefont
  {{Rees}}},\ }\href@noop {} {\bibfield  {journal} {\bibinfo  {journal} {The
  Observatory}\ }\textbf {\bibinfo {volume} {98}},\ \bibinfo {pages} {210}
  (\bibinfo {year} {1978})}\BibitemShut {NoStop}%
\bibitem [{\citenamefont {{Bahcall}}\ and\ \citenamefont
  {{Ostriker}}(1975)}]{Bahcall1975}%
  \BibitemOpen
  \bibfield  {author} {\bibinfo {author} {\bibfnamefont {J.~N.}\ \bibnamefont
  {{Bahcall}}}\ and\ \bibinfo {author} {\bibfnamefont {J.~P.}\ \bibnamefont
  {{Ostriker}}},\ }\href {\doibase 10.1038/256023a0} {\bibfield  {journal}
  {\bibinfo  {journal} {\nat}\ }\textbf {\bibinfo {volume} {256}},\ \bibinfo
  {pages} {23} (\bibinfo {year} {1975})}\BibitemShut {NoStop}%
\bibitem [{\citenamefont {{Begelman}}\ and\ \citenamefont
  {{Rees}}(1978)}]{Begelman1978}%
  \BibitemOpen
  \bibfield  {author} {\bibinfo {author} {\bibfnamefont {M.~C.}\ \bibnamefont
  {{Begelman}}}\ and\ \bibinfo {author} {\bibfnamefont {M.~J.}\ \bibnamefont
  {{Rees}}},\ }\href {\doibase 10.1093/mnras/185.4.847} {\bibfield  {journal}
  {\bibinfo  {journal} {mnras}\ }\textbf {\bibinfo {volume} {185}},\ \bibinfo
  {pages} {847} (\bibinfo {year} {1978})}\BibitemShut {NoStop}%
\bibitem [{\citenamefont {{Quinlan}}\ and\ \citenamefont
  {{Shapiro}}(1990)}]{Quinlan1990}%
  \BibitemOpen
  \bibfield  {author} {\bibinfo {author} {\bibfnamefont {G.~D.}\ \bibnamefont
  {{Quinlan}}}\ and\ \bibinfo {author} {\bibfnamefont {S.~L.}\ \bibnamefont
  {{Shapiro}}},\ }\href {\doibase 10.1086/168856} {\bibfield  {journal}
  {\bibinfo  {journal} {\apj}\ }\textbf {\bibinfo {volume} {356}},\ \bibinfo
  {pages} {483} (\bibinfo {year} {1990})}\BibitemShut {NoStop}%
\bibitem [{\citenamefont {Greene}\ \emph {et~al.}(2019)\citenamefont {Greene},
  \citenamefont {Strader},\ and\ \citenamefont {Ho}}]{Greene_Review}%
  \BibitemOpen
  \bibfield  {author} {\bibinfo {author} {\bibfnamefont {J.~E.}\ \bibnamefont
  {Greene}}, \bibinfo {author} {\bibfnamefont {J.}~\bibnamefont {Strader}}, \
  and\ \bibinfo {author} {\bibfnamefont {L.~C.}\ \bibnamefont {Ho}},\
  }\href@noop {} {\  (\bibinfo {year} {2019})},\ \Eprint
  {http://arxiv.org/abs/1911.09678} {arXiv:1911.09678 [astro-ph.GA]}
  \BibitemShut {NoStop}%
\bibitem [{\citenamefont {{Mezcua}}(2017)}]{Merzcua:2017}%
  \BibitemOpen
  \bibfield  {author} {\bibinfo {author} {\bibfnamefont {M.}~\bibnamefont
  {{Mezcua}}},\ }\href {\doibase 10.1142/S021827181730021X} {\bibfield
  {journal} {\bibinfo  {journal} {International Journal of Modern Physics D}\
  }\textbf {\bibinfo {volume} {26}},\ \bibinfo {eid} {1730021} (\bibinfo {year}
  {2017})},\ \Eprint {http://arxiv.org/abs/1705.09667} {arXiv:1705.09667}
  \BibitemShut {NoStop}%
\bibitem [{\citenamefont {{Koliopanos}}(2017)}]{2017mbhe.confE..51K}%
  \BibitemOpen
  \bibfield  {author} {\bibinfo {author} {\bibfnamefont {F.}~\bibnamefont
  {{Koliopanos}}},\ }in\ \href@noop {} {\emph {\bibinfo {booktitle}
  {Proceedings of the XII Multifrequency Behaviour of High Energy Cosmic
  Sources Workshop. 12-17 June}}}\ (\bibinfo {year} {2017})\ p.~\bibinfo
  {pages} {51},\ \Eprint {http://arxiv.org/abs/1801.01095} {arXiv:1801.01095
  [astro-ph.GA]} \BibitemShut {NoStop}%
\bibitem [{\citenamefont {Abbott}\ \emph
  {et~al.}(2019{\natexlab{b}})\citenamefont {Abbott} \emph
  {et~al.}}]{O1O2IMBHBPaper}%
  \BibitemOpen
  \bibfield  {author} {\bibinfo {author} {\bibfnamefont {B.~P.}\ \bibnamefont
  {Abbott}} \emph {et~al.} (\bibinfo {collaboration} {LIGO Scientific,
  Virgo}),\ }\href {\doibase 10.1103/PhysRevD.100.064064} {\bibfield  {journal}
  {\bibinfo  {journal} {Phys. Rev.}\ }\textbf {\bibinfo {volume} {D100}},\
  \bibinfo {pages} {064064} (\bibinfo {year} {2019}{\natexlab{b}})},\ \Eprint
  {http://arxiv.org/abs/1906.08000} {arXiv:1906.08000 [gr-qc]} \BibitemShut
  {NoStop}%
\bibitem [{\citenamefont {Mapelli}(2016)}]{Mapelli:2016vca}%
  \BibitemOpen
  \bibfield  {author} {\bibinfo {author} {\bibfnamefont {M.}~\bibnamefont
  {Mapelli}},\ }\href {\doibase 10.1093/mnras/stw869} {\bibfield  {journal}
  {\bibinfo  {journal} {Mon. Not. Roy. Astron. Soc.}\ }\textbf {\bibinfo
  {volume} {459}},\ \bibinfo {pages} {3432} (\bibinfo {year} {2016})},\ \Eprint
  {http://arxiv.org/abs/1604.03559} {arXiv:1604.03559 [astro-ph.GA]}
  \BibitemShut {NoStop}%
\bibitem [{\citenamefont {Calderon~Bustillo}\ \emph {et~al.}(2019)\citenamefont
  {Calderon~Bustillo}, \citenamefont {Sanchis-Gual}, \citenamefont
  {Torres-Forn\'e},\ and\ \citenamefont {Font}}]{HeadON}%
  \BibitemOpen
  \bibfield  {author} {\bibinfo {author} {\bibfnamefont {J.}~\bibnamefont
  {Calderon~Bustillo}}, \bibinfo {author} {\bibfnamefont {N.}~\bibnamefont
  {Sanchis-Gual}}, \bibinfo {author} {\bibfnamefont {A.}~\bibnamefont
  {Torres-Forn\'e}}, \ and\ \bibinfo {author} {\bibfnamefont {J.~A.}\
  \bibnamefont {Font}},\ }\href@noop {} {\bibfield  {journal} {\bibinfo
  {journal} {In prep.}\ } (\bibinfo {year} {2019})}\BibitemShut {NoStop}%
\bibitem [{\citenamefont {Rodriguez}\ \emph {et~al.}(2016)\citenamefont
  {Rodriguez}, \citenamefont {Chatterjee},\ and\ \citenamefont
  {Rasio}}]{Rodriguez:2016kxx}%
  \BibitemOpen
  \bibfield  {author} {\bibinfo {author} {\bibfnamefont {C.~L.}\ \bibnamefont
  {Rodriguez}}, \bibinfo {author} {\bibfnamefont {S.}~\bibnamefont
  {Chatterjee}}, \ and\ \bibinfo {author} {\bibfnamefont {F.~A.}\ \bibnamefont
  {Rasio}},\ }\href {\doibase 10.1103/PhysRevD.93.084029} {\bibfield  {journal}
  {\bibinfo  {journal} {Phys. Rev.}\ }\textbf {\bibinfo {volume} {D93}},\
  \bibinfo {pages} {084029} (\bibinfo {year} {2016})},\ \Eprint
  {http://arxiv.org/abs/1602.02444} {arXiv:1602.02444 [astro-ph.HE]}
  \BibitemShut {NoStop}%
\bibitem [{\citenamefont {Dal~Canton}\ \emph
  {et~al.}(2014{\natexlab{a}})\citenamefont {Dal~Canton} \emph
  {et~al.}}]{Canton:2014ena}%
  \BibitemOpen
  \bibfield  {author} {\bibinfo {author} {\bibfnamefont {T.}~\bibnamefont
  {Dal~Canton}} \emph {et~al.},\ }\href {\doibase 10.1103/PhysRevD.90.082004}
  {\bibfield  {journal} {\bibinfo  {journal} {Phys. Rev.}\ }\textbf {\bibinfo
  {volume} {D90}},\ \bibinfo {pages} {082004} (\bibinfo {year}
  {2014}{\natexlab{a}})},\ \Eprint {http://arxiv.org/abs/1405.6731}
  {arXiv:1405.6731 [gr-qc]} \BibitemShut {NoStop}%
\bibitem [{\citenamefont {Usman}\ \emph {et~al.}(2016)\citenamefont {Usman}
  \emph {et~al.}}]{Usman:2015kfa}%
  \BibitemOpen
  \bibfield  {author} {\bibinfo {author} {\bibfnamefont {S.~A.}\ \bibnamefont
  {Usman}} \emph {et~al.},\ }\href {\doibase 10.1088/0264-9381/33/21/215004}
  {\bibfield  {journal} {\bibinfo  {journal} {Class. Quant. Grav.}\ }\textbf
  {\bibinfo {volume} {33}},\ \bibinfo {pages} {215004} (\bibinfo {year}
  {2016})},\ \Eprint {http://arxiv.org/abs/1508.02357} {arXiv:1508.02357
  [gr-qc]} \BibitemShut {NoStop}%
\bibitem [{\citenamefont {Nitz}\ \emph
  {et~al.}(2017{\natexlab{a}})\citenamefont {Nitz}, \citenamefont {Harry},
  \citenamefont {Willis}, \citenamefont {Biwer}, \citenamefont {Brown},
  \citenamefont {Pekowsky}, \citenamefont {Dal~Canton}, \citenamefont
  {Williamson}, \citenamefont {Dent}, \citenamefont {Capano}, \citenamefont
  {Massinger}, \citenamefont {Lenon}, \citenamefont {Nielsen},\ and\
  \citenamefont {Cabero}}]{pycbc-github}%
  \BibitemOpen
  \bibfield  {author} {\bibinfo {author} {\bibfnamefont {A.~H.}\ \bibnamefont
  {Nitz}}, \bibinfo {author} {\bibfnamefont {I.~W.}\ \bibnamefont {Harry}},
  \bibinfo {author} {\bibfnamefont {J.~L.}\ \bibnamefont {Willis}}, \bibinfo
  {author} {\bibfnamefont {C.~M.}\ \bibnamefont {Biwer}}, \bibinfo {author}
  {\bibfnamefont {D.~A.}\ \bibnamefont {Brown}}, \bibinfo {author}
  {\bibfnamefont {L.~P.}\ \bibnamefont {Pekowsky}}, \bibinfo {author}
  {\bibfnamefont {T.}~\bibnamefont {Dal~Canton}}, \bibinfo {author}
  {\bibfnamefont {A.~R.}\ \bibnamefont {Williamson}}, \bibinfo {author}
  {\bibfnamefont {T.}~\bibnamefont {Dent}}, \bibinfo {author} {\bibfnamefont
  {C.~D.}\ \bibnamefont {Capano}}, \bibinfo {author} {\bibfnamefont {T.~J.}\
  \bibnamefont {Massinger}}, \bibinfo {author} {\bibfnamefont {A.~K.}\
  \bibnamefont {Lenon}}, \bibinfo {author} {\bibfnamefont {A.~B.}\ \bibnamefont
  {Nielsen}}, \ and\ \bibinfo {author} {\bibfnamefont {M.}~\bibnamefont
  {Cabero}},\ }\href {\doibase 10.5281/zenodo.344823} {\enquote {\bibinfo
  {title} {{PyCBC Software}},}\ }\bibinfo {howpublished}
  {\href{https://github.com/ligo-cbc/pycbc}{github.com/ligo-cbc/pycbc}}
  (\bibinfo {year} {2017}{\natexlab{a}})\BibitemShut {NoStop}%
\bibitem [{\citenamefont {Klimenko}\ \emph {et~al.}(2016)\citenamefont
  {Klimenko} \emph {et~al.}}]{Klimenko:2015ypf}%
  \BibitemOpen
  \bibfield  {author} {\bibinfo {author} {\bibfnamefont {S.}~\bibnamefont
  {Klimenko}} \emph {et~al.},\ }\href {\doibase 10.1103/PhysRevD.93.042004}
  {\bibfield  {journal} {\bibinfo  {journal} {Phys. Rev.}\ }\textbf {\bibinfo
  {volume} {D93}},\ \bibinfo {pages} {042004} (\bibinfo {year} {2016})},\
  \Eprint {http://arxiv.org/abs/1511.05999} {arXiv:1511.05999 [gr-qc]}
  \BibitemShut {NoStop}%
\bibitem [{\citenamefont {Calder\'on~Bustillo}\ \emph
  {et~al.}(2017)\citenamefont {Calder\'on~Bustillo}, \citenamefont {Laguna},\
  and\ \citenamefont {Shoemaker}}]{CalderonBustillo:2017skv}%
  \BibitemOpen
  \bibfield  {author} {\bibinfo {author} {\bibfnamefont {J.}~\bibnamefont
  {Calder\'on~Bustillo}}, \bibinfo {author} {\bibfnamefont {P.}~\bibnamefont
  {Laguna}}, \ and\ \bibinfo {author} {\bibfnamefont {D.}~\bibnamefont
  {Shoemaker}},\ }\href {\doibase 10.1103/PhysRevD.95.104038} {\bibfield
  {journal} {\bibinfo  {journal} {Phys. Rev. D}\ }\textbf {\bibinfo {volume}
  {95}},\ \bibinfo {pages} {104038} (\bibinfo {year} {2017})}\BibitemShut
  {NoStop}%
\bibitem [{\citenamefont {Abbott}\ \emph
  {et~al.}(2019{\natexlab{c}})\citenamefont {Abbott} \emph
  {et~al.}}]{OpenData}%
  \BibitemOpen
  \bibfield  {author} {\bibinfo {author} {\bibfnamefont {R.}~\bibnamefont
  {Abbott}} \emph {et~al.} (\bibinfo {collaboration} {LIGO Scientific,
  Virgo}),\ }\href@noop {} {\  (\bibinfo {year} {2019}{\natexlab{c}})},\
  \Eprint {http://arxiv.org/abs/1912.11716} {arXiv:1912.11716 [gr-qc]}
  \BibitemShut {NoStop}%
\bibitem [{\citenamefont {Goldberg}\ \emph {et~al.}(1967)\citenamefont
  {Goldberg}, \citenamefont {MacFarlane}, \citenamefont {Newman}, \citenamefont
  {Rohrlich},\ and\ \citenamefont {Sudarshan}}]{Goldberg:1966uu}%
  \BibitemOpen
  \bibfield  {author} {\bibinfo {author} {\bibfnamefont {J.~N.}\ \bibnamefont
  {Goldberg}}, \bibinfo {author} {\bibfnamefont {A.~J.}\ \bibnamefont
  {MacFarlane}}, \bibinfo {author} {\bibfnamefont {E.~T.}\ \bibnamefont
  {Newman}}, \bibinfo {author} {\bibfnamefont {F.}~\bibnamefont {Rohrlich}}, \
  and\ \bibinfo {author} {\bibfnamefont {E.~C.~G.}\ \bibnamefont {Sudarshan}},\
  }\href {\doibase 10.1063/1.1705135} {\bibfield  {journal} {\bibinfo
  {journal} {J. Math. Phys.}\ }\textbf {\bibinfo {volume} {8}},\ \bibinfo
  {pages} {2155} (\bibinfo {year} {1967})}\BibitemShut {NoStop}%
\bibitem [{\citenamefont {Maggiore}(2008)}]{maggiore2008gravitational}%
  \BibitemOpen
  \bibfield  {author} {\bibinfo {author} {\bibfnamefont {M.}~\bibnamefont
  {Maggiore}},\ }\href {https://books.google.co.in/books?id=AqVpQgAACAAJ}
  {\emph {\bibinfo {title} {Gravitational Waves: Volume 1: Theory and
  Experiments}}},\ Gravitational Waves\ (\bibinfo  {publisher} {OUP Oxford},\
  \bibinfo {year} {2008})\BibitemShut {NoStop}%
\bibitem [{\citenamefont {Creighton}\ and\ \citenamefont
  {Anderson}(2012)}]{GW-BOOK-Anderson}%
  \BibitemOpen
  \bibfield  {author} {\bibinfo {author} {\bibfnamefont {J.}~\bibnamefont
  {Creighton}}\ and\ \bibinfo {author} {\bibfnamefont {W.}~\bibnamefont
  {Anderson}},\ }\href {https://books.google.co.in/books?id=W\_TVS\_6JYJcC}
  {\emph {\bibinfo {title} {Gravitational-Wave Physics and Astronomy: An
  Introduction to Theory, Experiment and Data Analysis}}},\ Wiley Series in
  Cosmology\ (\bibinfo  {publisher} {Wiley},\ \bibinfo {year}
  {2012})\BibitemShut {NoStop}%
\bibitem [{\citenamefont {Pekowsky}\ \emph {et~al.}(2013)\citenamefont
  {Pekowsky}, \citenamefont {Healy}, \citenamefont {Shoemaker},\ and\
  \citenamefont {Laguna}}]{Pekowsky:2012sr}%
  \BibitemOpen
  \bibfield  {author} {\bibinfo {author} {\bibfnamefont {L.}~\bibnamefont
  {Pekowsky}}, \bibinfo {author} {\bibfnamefont {J.}~\bibnamefont {Healy}},
  \bibinfo {author} {\bibfnamefont {D.}~\bibnamefont {Shoemaker}}, \ and\
  \bibinfo {author} {\bibfnamefont {P.}~\bibnamefont {Laguna}},\ }\href
  {\doibase 10.1103/PhysRevD.87.084008} {\bibfield  {journal} {\bibinfo
  {journal} {Phys.Rev.}\ }\textbf {\bibinfo {volume} {D87}},\ \bibinfo {pages}
  {084008} (\bibinfo {year} {2013})},\ \Eprint {http://arxiv.org/abs/1210.1891}
  {arXiv:1210.1891 [gr-qc]} \BibitemShut {NoStop}%
\bibitem [{\citenamefont {Varma}\ \emph {et~al.}(2014)\citenamefont {Varma},
  \citenamefont {Ajith}, \citenamefont {Husa}, \citenamefont {Bustillo},
  \citenamefont {Hannam},\ and\ \citenamefont {P\"urrer}}]{Varma:2014jxa}%
  \BibitemOpen
  \bibfield  {author} {\bibinfo {author} {\bibfnamefont {V.}~\bibnamefont
  {Varma}}, \bibinfo {author} {\bibfnamefont {P.}~\bibnamefont {Ajith}},
  \bibinfo {author} {\bibfnamefont {S.}~\bibnamefont {Husa}}, \bibinfo {author}
  {\bibfnamefont {J.~C.}\ \bibnamefont {Bustillo}}, \bibinfo {author}
  {\bibfnamefont {M.}~\bibnamefont {Hannam}}, \ and\ \bibinfo {author}
  {\bibfnamefont {M.}~\bibnamefont {P\"urrer}},\ }\href {\doibase
  10.1103/PhysRevD.90.124004} {\bibfield  {journal} {\bibinfo  {journal} {Phys.
  Rev.}\ }\textbf {\bibinfo {volume} {D90}},\ \bibinfo {pages} {124004}
  (\bibinfo {year} {2014})},\ \Eprint {http://arxiv.org/abs/1409.2349}
  {arXiv:1409.2349 [gr-qc]} \BibitemShut {NoStop}%
\bibitem [{\citenamefont {Calder{\'o}n~Bustillo}\ \emph
  {et~al.}(2016)\citenamefont {Calder{\'o}n~Bustillo}, \citenamefont {Husa},
  \citenamefont {Sintes},\ and\ \citenamefont {P{\"u}rrer}}]{Bustillo:2015qty}%
  \BibitemOpen
  \bibfield  {author} {\bibinfo {author} {\bibfnamefont {J.}~\bibnamefont
  {Calder{\'o}n~Bustillo}}, \bibinfo {author} {\bibfnamefont {S.}~\bibnamefont
  {Husa}}, \bibinfo {author} {\bibfnamefont {A.~M.}\ \bibnamefont {Sintes}}, \
  and\ \bibinfo {author} {\bibfnamefont {M.}~\bibnamefont {P{\"u}rrer}},\
  }\href {\doibase 10.1103/PhysRevD.93.084019} {\bibfield  {journal} {\bibinfo
  {journal} {Phys. Rev.}\ }\textbf {\bibinfo {volume} {D93}},\ \bibinfo {pages}
  {084019} (\bibinfo {year} {2016})},\ \Eprint
  {http://arxiv.org/abs/1511.02060} {arXiv:1511.02060 [gr-qc]} \BibitemShut
  {NoStop}%
\bibitem [{\citenamefont {Varma}\ and\ \citenamefont
  {Ajith}(2017)}]{Varma:2016dnf}%
  \BibitemOpen
  \bibfield  {author} {\bibinfo {author} {\bibfnamefont {V.}~\bibnamefont
  {Varma}}\ and\ \bibinfo {author} {\bibfnamefont {P.}~\bibnamefont {Ajith}},\
  }\href {\doibase 10.1103/PhysRevD.96.124024} {\bibfield  {journal} {\bibinfo
  {journal} {Phys. Rev. D}\ }\textbf {\bibinfo {volume} {96}},\ \bibinfo
  {pages} {124024} (\bibinfo {year} {2017})}\BibitemShut {NoStop}%
\bibitem [{\citenamefont {Calder{\'o}n~Bustillo}\ \emph
  {et~al.}(2017)\citenamefont {Calder{\'o}n~Bustillo}, \citenamefont {Laguna},\
  and\ \citenamefont {Shoemaker}}]{Bustillo:2016gid}%
  \BibitemOpen
  \bibfield  {author} {\bibinfo {author} {\bibfnamefont {J.}~\bibnamefont
  {Calder{\'o}n~Bustillo}}, \bibinfo {author} {\bibfnamefont {P.}~\bibnamefont
  {Laguna}}, \ and\ \bibinfo {author} {\bibfnamefont {D.}~\bibnamefont
  {Shoemaker}},\ }\href {\doibase 10.1103/PhysRevD.95.104038} {\bibfield
  {journal} {\bibinfo  {journal} {Phys. Rev.}\ }\textbf {\bibinfo {volume}
  {D95}},\ \bibinfo {pages} {104038} (\bibinfo {year} {2017})},\ \Eprint
  {http://arxiv.org/abs/1612.02340} {arXiv:1612.02340 [gr-qc]} \BibitemShut
  {NoStop}%
\bibitem [{\citenamefont {Graff}\ \emph {et~al.}(2015)\citenamefont {Graff},
  \citenamefont {Buonanno},\ and\ \citenamefont
  {Sathyaprakash}}]{Graff:2015bba}%
  \BibitemOpen
  \bibfield  {author} {\bibinfo {author} {\bibfnamefont {P.~B.}\ \bibnamefont
  {Graff}}, \bibinfo {author} {\bibfnamefont {A.}~\bibnamefont {Buonanno}}, \
  and\ \bibinfo {author} {\bibfnamefont {B.}~\bibnamefont {Sathyaprakash}},\
  }\href {\doibase 10.1103/PhysRevD.92.022002} {\bibfield  {journal} {\bibinfo
  {journal} {Phys. Rev.}\ }\textbf {\bibinfo {volume} {D92}},\ \bibinfo {pages}
  {022002} (\bibinfo {year} {2015})},\ \Eprint
  {http://arxiv.org/abs/1504.04766} {arXiv:1504.04766 [gr-qc]} \BibitemShut
  {NoStop}%
\bibitem [{\citenamefont {Calder{\'o}n~Bustillo}\ \emph
  {et~al.}(2018)\citenamefont {Calder{\'o}n~Bustillo}, \citenamefont {Clark},
  \citenamefont {Laguna},\ and\ \citenamefont
  {Shoemaker}}]{CalderonBustillo:2018zuq}%
  \BibitemOpen
  \bibfield  {author} {\bibinfo {author} {\bibfnamefont {J.}~\bibnamefont
  {Calder{\'o}n~Bustillo}}, \bibinfo {author} {\bibfnamefont {J.~A.}\
  \bibnamefont {Clark}}, \bibinfo {author} {\bibfnamefont {P.}~\bibnamefont
  {Laguna}}, \ and\ \bibinfo {author} {\bibfnamefont {D.}~\bibnamefont
  {Shoemaker}},\ }\href {\doibase 10.1103/PhysRevLett.121.191102} {\bibfield
  {journal} {\bibinfo  {journal} {Phys. Rev. Lett.}\ }\textbf {\bibinfo
  {volume} {121}},\ \bibinfo {pages} {191102} (\bibinfo {year} {2018})},\
  \Eprint {http://arxiv.org/abs/1806.11160} {arXiv:1806.11160 [gr-qc]}
  \BibitemShut {NoStop}%
\bibitem [{\citenamefont {{Calder{\'o}n Bustillo}}\ \emph
  {et~al.}(2019)\citenamefont {{Calder{\'o}n Bustillo}}, \citenamefont
  {{Evans}}, \citenamefont {{Clark}}, \citenamefont {{Kim}}, \citenamefont
  {{Laguna}},\ and\ \citenamefont {{Shoemaker}}}]{CalderonBustillo:2019wwe}%
  \BibitemOpen
  \bibfield  {author} {\bibinfo {author} {\bibfnamefont {J.}~\bibnamefont
  {{Calder{\'o}n Bustillo}}}, \bibinfo {author} {\bibfnamefont
  {C.}~\bibnamefont {{Evans}}}, \bibinfo {author} {\bibfnamefont {J.~A.}\
  \bibnamefont {{Clark}}}, \bibinfo {author} {\bibfnamefont {G.}~\bibnamefont
  {{Kim}}}, \bibinfo {author} {\bibfnamefont {P.}~\bibnamefont {{Laguna}}}, \
  and\ \bibinfo {author} {\bibfnamefont {D.}~\bibnamefont {{Shoemaker}}},\
  }\href@noop {} {\bibfield  {journal} {\bibinfo  {journal} {arXiv e-prints}\
  ,\ \bibinfo {eid} {arXiv:1906.01153}} (\bibinfo {year} {2019})},\ \Eprint
  {http://arxiv.org/abs/1906.01153} {arXiv:1906.01153 [gr-qc]} \BibitemShut
  {NoStop}%
\bibitem [{\citenamefont {Poisson}\ and\ \citenamefont
  {Will}(1995)}]{Will1995}%
  \BibitemOpen
  \bibfield  {author} {\bibinfo {author} {\bibfnamefont {E.}~\bibnamefont
  {Poisson}}\ and\ \bibinfo {author} {\bibfnamefont {C.~M.}\ \bibnamefont
  {Will}},\ }\href {\doibase 10.1103/PhysRevD.52.848} {\bibfield  {journal}
  {\bibinfo  {journal} {Phys. Rev. D}\ }\textbf {\bibinfo {volume} {52}},\
  \bibinfo {pages} {848} (\bibinfo {year} {1995})}\BibitemShut {NoStop}%
\bibitem [{\citenamefont {Santamaria}\ \emph {et~al.}(2010)\citenamefont
  {Santamaria} \emph {et~al.}}]{Santamaria:2010yb}%
  \BibitemOpen
  \bibfield  {author} {\bibinfo {author} {\bibfnamefont {L.}~\bibnamefont
  {Santamaria}} \emph {et~al.},\ }\href {\doibase 10.1103/PhysRevD.82.064016}
  {\bibfield  {journal} {\bibinfo  {journal} {Phys. Rev.}\ }\textbf {\bibinfo
  {volume} {D82}},\ \bibinfo {pages} {064016} (\bibinfo {year} {2010})},\
  \Eprint {http://arxiv.org/abs/1005.3306} {arXiv:1005.3306 [gr-qc]}
  \BibitemShut {NoStop}%
\bibitem [{\citenamefont {Schmidt}\ \emph {et~al.}(2015)\citenamefont
  {Schmidt}, \citenamefont {Ohme},\ and\ \citenamefont
  {Hannam}}]{Schmidt:2014iyl}%
  \BibitemOpen
  \bibfield  {author} {\bibinfo {author} {\bibfnamefont {P.}~\bibnamefont
  {Schmidt}}, \bibinfo {author} {\bibfnamefont {F.}~\bibnamefont {Ohme}}, \
  and\ \bibinfo {author} {\bibfnamefont {M.}~\bibnamefont {Hannam}},\ }\href
  {\doibase 10.1103/PhysRevD.91.024043} {\bibfield  {journal} {\bibinfo
  {journal} {Phys. Rev.}\ }\textbf {\bibinfo {volume} {D91}},\ \bibinfo {pages}
  {024043} (\bibinfo {year} {2015})},\ \Eprint {http://arxiv.org/abs/1408.1810}
  {arXiv:1408.1810 [gr-qc]} \BibitemShut {NoStop}%
\bibitem [{\citenamefont {Gayathri}\ \emph {et~al.}(2019)\citenamefont
  {Gayathri}, \citenamefont {Bacon}, \citenamefont {Pai}, \citenamefont
  {Chassande-Mottin}, \citenamefont {Salemi},\ and\ \citenamefont
  {Vedovato}}]{Gayathri:2019omo}%
  \BibitemOpen
  \bibfield  {author} {\bibinfo {author} {\bibfnamefont {V.}~\bibnamefont
  {Gayathri}}, \bibinfo {author} {\bibfnamefont {P.}~\bibnamefont {Bacon}},
  \bibinfo {author} {\bibfnamefont {A.}~\bibnamefont {Pai}}, \bibinfo {author}
  {\bibfnamefont {E.}~\bibnamefont {Chassande-Mottin}}, \bibinfo {author}
  {\bibfnamefont {F.}~\bibnamefont {Salemi}}, \ and\ \bibinfo {author}
  {\bibfnamefont {G.}~\bibnamefont {Vedovato}},\ }\href {\doibase
  10.1103/PhysRevD.100.124022} {\bibfield  {journal} {\bibinfo  {journal}
  {Phys. Rev. D}\ }\textbf {\bibinfo {volume} {100}},\ \bibinfo {pages}
  {124022} (\bibinfo {year} {2019})}\BibitemShut {NoStop}%
\bibitem [{\citenamefont {Tiwari}\ \emph {et~al.}(2016)\citenamefont {Tiwari}
  \emph {et~al.}}]{Tiwari:2015gal}%
  \BibitemOpen
  \bibfield  {author} {\bibinfo {author} {\bibfnamefont {V.}~\bibnamefont
  {Tiwari}} \emph {et~al.},\ }\href {\doibase 10.1103/PhysRevD.93.043007}
  {\bibfield  {journal} {\bibinfo  {journal} {Phys. Rev.}\ }\textbf {\bibinfo
  {volume} {D93}},\ \bibinfo {pages} {043007} (\bibinfo {year} {2016})},\
  \Eprint {http://arxiv.org/abs/1511.09240} {arXiv:1511.09240 [gr-qc]}
  \BibitemShut {NoStop}%
\bibitem [{\citenamefont {Owen}\ and\ \citenamefont
  {Sathyaprakash}(1999)}]{Owen:1998dk}%
  \BibitemOpen
  \bibfield  {author} {\bibinfo {author} {\bibfnamefont {B.~J.}\ \bibnamefont
  {Owen}}\ and\ \bibinfo {author} {\bibfnamefont {B.~S.}\ \bibnamefont
  {Sathyaprakash}},\ }\href {\doibase 10.1103/PhysRevD.60.022002} {\bibfield
  {journal} {\bibinfo  {journal} {Phys. Rev.}\ }\textbf {\bibinfo {volume}
  {D60}},\ \bibinfo {pages} {022002} (\bibinfo {year} {1999})},\ \Eprint
  {http://arxiv.org/abs/gr-qc/9808076} {arXiv:gr-qc/9808076 [gr-qc]}
  \BibitemShut {NoStop}%
\bibitem [{\citenamefont {Vainshtein}\ and\ \citenamefont
  {Zubakov}(1970)}]{1970esn..book.....W}%
  \BibitemOpen
  \bibfield  {author} {\bibinfo {author} {\bibfnamefont {L.}~\bibnamefont
  {Vainshtein}}\ and\ \bibinfo {author} {\bibfnamefont {V.}~\bibnamefont
  {Zubakov}},\ }\href {https://books.google.co.in/books?id=PTPSzAEACAAJ} {\emph
  {\bibinfo {title} {Extraction of Signals from Noise: By L.A. Wainstein and
  V.D. Zubakov}}}\ (\bibinfo  {publisher} {Dover},\ \bibinfo {year}
  {1970})\BibitemShut {NoStop}%
\bibitem [{\citenamefont {Dal~Canton}\ \emph
  {et~al.}(2014{\natexlab{b}})\citenamefont {Dal~Canton}, \citenamefont
  {Bhagwat}, \citenamefont {Dhurandhar},\ and\ \citenamefont
  {Lundgren}}]{Canton:2013joa}%
  \BibitemOpen
  \bibfield  {author} {\bibinfo {author} {\bibfnamefont {T.}~\bibnamefont
  {Dal~Canton}}, \bibinfo {author} {\bibfnamefont {S.}~\bibnamefont {Bhagwat}},
  \bibinfo {author} {\bibfnamefont {S.}~\bibnamefont {Dhurandhar}}, \ and\
  \bibinfo {author} {\bibfnamefont {A.}~\bibnamefont {Lundgren}},\ }\href
  {\doibase 10.1088/0264-9381/31/1/015016} {\bibfield  {journal} {\bibinfo
  {journal} {Class.Quant.Grav.}\ }\textbf {\bibinfo {volume} {31}},\ \bibinfo
  {pages} {015016} (\bibinfo {year} {2014}{\natexlab{b}})},\ \Eprint
  {http://arxiv.org/abs/1304.0008} {arXiv:1304.0008 [gr-qc]} \BibitemShut
  {NoStop}%
\bibitem [{\citenamefont {Nitz}(2018)}]{Nitz:2017lco}%
  \BibitemOpen
  \bibfield  {author} {\bibinfo {author} {\bibfnamefont {A.~H.}\ \bibnamefont
  {Nitz}},\ }\href {\doibase 10.1088/1361-6382/aaa13d} {\bibfield  {journal}
  {\bibinfo  {journal} {Class. Quant. Grav.}\ }\textbf {\bibinfo {volume}
  {35}},\ \bibinfo {pages} {035016} (\bibinfo {year} {2018})},\ \Eprint
  {http://arxiv.org/abs/1709.08974} {arXiv:1709.08974 [gr-qc]} \BibitemShut
  {NoStop}%
\bibitem [{\citenamefont {Messick}\ \emph {et~al.}(2017)\citenamefont {Messick}
  \emph {et~al.}}]{Messick:2016aqy}%
  \BibitemOpen
  \bibfield  {author} {\bibinfo {author} {\bibfnamefont {C.}~\bibnamefont
  {Messick}} \emph {et~al.},\ }\href {\doibase 10.1103/PhysRevD.95.042001}
  {\bibfield  {journal} {\bibinfo  {journal} {Phys. Rev.}\ }\textbf {\bibinfo
  {volume} {D95}},\ \bibinfo {pages} {042001} (\bibinfo {year} {2017})},\
  \Eprint {http://arxiv.org/abs/1604.04324} {arXiv:1604.04324 [astro-ph.IM]}
  \BibitemShut {NoStop}%
\bibitem [{\citenamefont {Allen}\ \emph {et~al.}(2012)\citenamefont {Allen},
  \citenamefont {Anderson}, \citenamefont {Brady}, \citenamefont {Brown},\ and\
  \citenamefont {Creighton}}]{Allen:2005fk}%
  \BibitemOpen
  \bibfield  {author} {\bibinfo {author} {\bibfnamefont {B.}~\bibnamefont
  {Allen}}, \bibinfo {author} {\bibfnamefont {W.~G.}\ \bibnamefont {Anderson}},
  \bibinfo {author} {\bibfnamefont {P.~R.}\ \bibnamefont {Brady}}, \bibinfo
  {author} {\bibfnamefont {D.~A.}\ \bibnamefont {Brown}}, \ and\ \bibinfo
  {author} {\bibfnamefont {J.~D.~E.}\ \bibnamefont {Creighton}},\ }\href
  {\doibase 10.1103/PhysRevD.85.122006} {\bibfield  {journal} {\bibinfo
  {journal} {Phys. Rev. D}\ }\textbf {\bibinfo {volume} {85}},\ \bibinfo
  {pages} {122006} (\bibinfo {year} {2012})}\BibitemShut {NoStop}%
\bibitem [{\citenamefont {Davies}\ \emph {et~al.}(2020)\citenamefont {Davies},
  \citenamefont {Dent}, \citenamefont {Tápai}, \citenamefont {Harry},
  \citenamefont {McIsaac},\ and\ \citenamefont {Nitz}}]{Davies:2020tsx}%
  \BibitemOpen
  \bibfield  {author} {\bibinfo {author} {\bibfnamefont {G.~S.}\ \bibnamefont
  {Davies}}, \bibinfo {author} {\bibfnamefont {T.}~\bibnamefont {Dent}},
  \bibinfo {author} {\bibfnamefont {M.}~\bibnamefont {Tápai}}, \bibinfo
  {author} {\bibfnamefont {I.}~\bibnamefont {Harry}}, \bibinfo {author}
  {\bibfnamefont {C.}~\bibnamefont {McIsaac}}, \ and\ \bibinfo {author}
  {\bibfnamefont {A.~H.}\ \bibnamefont {Nitz}},\ }\href@noop {} {\  (\bibinfo
  {year} {2020})},\ \Eprint {http://arxiv.org/abs/2002.08291} {arXiv:2002.08291
  [astro-ph.HE]} \BibitemShut {NoStop}%
\bibitem [{\citenamefont {Babak}\ \emph {et~al.}(2013)\citenamefont {Babak},
  \citenamefont {Biswas}, \citenamefont {Brady}, \citenamefont {Brown},
  \citenamefont {Cannon}, \citenamefont {Capano}, \citenamefont {Clayton},
  \citenamefont {Cokelaer}, \citenamefont {Creighton}, \citenamefont {Dent},
  \citenamefont {Dietz}, \citenamefont {Fairhurst}, \citenamefont {Fotopoulos},
  \citenamefont {Gonz\'alez}, \citenamefont {Hanna}, \citenamefont {Harry},
  \citenamefont {Jones}, \citenamefont {Keppel}, \citenamefont {McKechan},
  \citenamefont {Pekowsky}, \citenamefont {Privitera}, \citenamefont
  {Robinson}, \citenamefont {Rodriguez}, \citenamefont {Sathyaprakash},
  \citenamefont {Sengupta}, \citenamefont {Vallisneri}, \citenamefont
  {Vaulin},\ and\ \citenamefont {Weinstein}}]{PhysRevD.87.024033}%
  \BibitemOpen
  \bibfield  {author} {\bibinfo {author} {\bibfnamefont {S.}~\bibnamefont
  {Babak}}, \bibinfo {author} {\bibfnamefont {R.}~\bibnamefont {Biswas}},
  \bibinfo {author} {\bibfnamefont {P.~R.}\ \bibnamefont {Brady}}, \bibinfo
  {author} {\bibfnamefont {D.~A.}\ \bibnamefont {Brown}}, \bibinfo {author}
  {\bibfnamefont {K.}~\bibnamefont {Cannon}}, \bibinfo {author} {\bibfnamefont
  {C.~D.}\ \bibnamefont {Capano}}, \bibinfo {author} {\bibfnamefont {J.~H.}\
  \bibnamefont {Clayton}}, \bibinfo {author} {\bibfnamefont {T.}~\bibnamefont
  {Cokelaer}}, \bibinfo {author} {\bibfnamefont {J.~D.~E.}\ \bibnamefont
  {Creighton}}, \bibinfo {author} {\bibfnamefont {T.}~\bibnamefont {Dent}},
  \bibinfo {author} {\bibfnamefont {A.}~\bibnamefont {Dietz}}, \bibinfo
  {author} {\bibfnamefont {S.}~\bibnamefont {Fairhurst}}, \bibinfo {author}
  {\bibfnamefont {N.}~\bibnamefont {Fotopoulos}}, \bibinfo {author}
  {\bibfnamefont {G.}~\bibnamefont {Gonz\'alez}}, \bibinfo {author}
  {\bibfnamefont {C.}~\bibnamefont {Hanna}}, \bibinfo {author} {\bibfnamefont
  {I.~W.}\ \bibnamefont {Harry}}, \bibinfo {author} {\bibfnamefont
  {G.}~\bibnamefont {Jones}}, \bibinfo {author} {\bibfnamefont
  {D.}~\bibnamefont {Keppel}}, \bibinfo {author} {\bibfnamefont {D.~J.~A.}\
  \bibnamefont {McKechan}}, \bibinfo {author} {\bibfnamefont {L.}~\bibnamefont
  {Pekowsky}}, \bibinfo {author} {\bibfnamefont {S.}~\bibnamefont {Privitera}},
  \bibinfo {author} {\bibfnamefont {C.}~\bibnamefont {Robinson}}, \bibinfo
  {author} {\bibfnamefont {A.~C.}\ \bibnamefont {Rodriguez}}, \bibinfo {author}
  {\bibfnamefont {B.~S.}\ \bibnamefont {Sathyaprakash}}, \bibinfo {author}
  {\bibfnamefont {A.~S.}\ \bibnamefont {Sengupta}}, \bibinfo {author}
  {\bibfnamefont {M.}~\bibnamefont {Vallisneri}}, \bibinfo {author}
  {\bibfnamefont {R.}~\bibnamefont {Vaulin}}, \ and\ \bibinfo {author}
  {\bibfnamefont {A.~J.}\ \bibnamefont {Weinstein}},\ }\href {\doibase
  10.1103/PhysRevD.87.024033} {\bibfield  {journal} {\bibinfo  {journal} {Phys.
  Rev. D}\ }\textbf {\bibinfo {volume} {87}},\ \bibinfo {pages} {024033}
  (\bibinfo {year} {2013})}\BibitemShut {NoStop}%
\bibitem [{\citenamefont {Nitz}\ \emph
  {et~al.}(2017{\natexlab{b}})\citenamefont {Nitz}, \citenamefont {Dent},
  \citenamefont {Dal~Canton}, \citenamefont {Fairhurst},\ and\ \citenamefont
  {Brown}}]{Nitz:2017svb}%
  \BibitemOpen
  \bibfield  {author} {\bibinfo {author} {\bibfnamefont {A.~H.}\ \bibnamefont
  {Nitz}}, \bibinfo {author} {\bibfnamefont {T.}~\bibnamefont {Dent}}, \bibinfo
  {author} {\bibfnamefont {T.}~\bibnamefont {Dal~Canton}}, \bibinfo {author}
  {\bibfnamefont {S.}~\bibnamefont {Fairhurst}}, \ and\ \bibinfo {author}
  {\bibfnamefont {D.~A.}\ \bibnamefont {Brown}},\ }\href {\doibase
  10.3847/1538-4357/aa8f50} {\bibfield  {journal} {\bibinfo  {journal}
  {Astrophys. J.}\ }\textbf {\bibinfo {volume} {849}},\ \bibinfo {pages} {118}
  (\bibinfo {year} {2017}{\natexlab{b}})},\ \Eprint
  {http://arxiv.org/abs/1705.01513} {arXiv:1705.01513 [gr-qc]} \BibitemShut
  {NoStop}%
\bibitem [{\citenamefont {{Dal Canton}}\ and\ \citenamefont
  {{Harry}}(2017)}]{DalCanton:2017ala}%
  \BibitemOpen
  \bibfield  {author} {\bibinfo {author} {\bibfnamefont {T.}~\bibnamefont {{Dal
  Canton}}}\ and\ \bibinfo {author} {\bibfnamefont {I.~W.}\ \bibnamefont
  {{Harry}}},\ }\href@noop {} {\bibfield  {journal} {\bibinfo  {journal} {arXiv
  e-prints}\ ,\ \bibinfo {eid} {arXiv:1705.01845}} (\bibinfo {year} {2017})},\
  \Eprint {http://arxiv.org/abs/1705.01845} {arXiv:1705.01845 [gr-qc]}
  \BibitemShut {NoStop}%
\bibitem [{\citenamefont {Canton}\ \emph {et~al.}(2013)\citenamefont {Canton},
  \citenamefont {Bhagwat}, \citenamefont {Dhurandhar},\ and\ \citenamefont
  {Lundgren}}]{Canton2013}%
  \BibitemOpen
  \bibfield  {author} {\bibinfo {author} {\bibfnamefont {T.~D.}\ \bibnamefont
  {Canton}}, \bibinfo {author} {\bibfnamefont {S.}~\bibnamefont {Bhagwat}},
  \bibinfo {author} {\bibfnamefont {S.~V.}\ \bibnamefont {Dhurandhar}}, \ and\
  \bibinfo {author} {\bibfnamefont {A.}~\bibnamefont {Lundgren}},\ }\href
  {\doibase 10.1088/0264-9381/31/1/015016} {\bibfield  {journal} {\bibinfo
  {journal} {Classical and Quantum Gravity}\ }\textbf {\bibinfo {volume}
  {31}},\ \bibinfo {pages} {015016} (\bibinfo {year} {2013})}\BibitemShut
  {NoStop}%
\bibitem [{\citenamefont {Boh{\'e}}\ \emph {et~al.}(2017)\citenamefont
  {Boh{\'e}} \emph {et~al.}}]{Bohe:2016gbl}%
  \BibitemOpen
  \bibfield  {author} {\bibinfo {author} {\bibfnamefont {A.}~\bibnamefont
  {Boh{\'e}}} \emph {et~al.},\ }\href {\doibase 10.1103/PhysRevD.95.044028}
  {\bibfield  {journal} {\bibinfo  {journal} {Phys. Rev.}\ }\textbf {\bibinfo
  {volume} {D95}},\ \bibinfo {pages} {044028} (\bibinfo {year} {2017})},\
  \Eprint {http://arxiv.org/abs/1611.03703} {arXiv:1611.03703 [gr-qc]}
  \BibitemShut {NoStop}%
\bibitem [{\citenamefont {{Jani}}\ \emph {et~al.}(2016)\citenamefont {{Jani}},
  \citenamefont {{Healy}}, \citenamefont {{Clark}}, \citenamefont {{London}},
  \citenamefont {{Laguna}},\ and\ \citenamefont
  {{Shoemaker}}}]{2016CQGra..33t4001J}%
  \BibitemOpen
  \bibfield  {author} {\bibinfo {author} {\bibfnamefont {K.}~\bibnamefont
  {{Jani}}}, \bibinfo {author} {\bibfnamefont {J.}~\bibnamefont {{Healy}}},
  \bibinfo {author} {\bibfnamefont {J.~A.}\ \bibnamefont {{Clark}}}, \bibinfo
  {author} {\bibfnamefont {L.}~\bibnamefont {{London}}}, \bibinfo {author}
  {\bibfnamefont {P.}~\bibnamefont {{Laguna}}}, \ and\ \bibinfo {author}
  {\bibfnamefont {D.}~\bibnamefont {{Shoemaker}}},\ }\href {\doibase
  10.1088/0264-9381/33/20/204001} {\bibfield  {journal} {\bibinfo  {journal}
  {Classical and Quantum Gravity}\ }\textbf {\bibinfo {volume} {33}},\ \bibinfo
  {eid} {204001} (\bibinfo {year} {2016})},\ \Eprint
  {http://arxiv.org/abs/1605.03204} {arXiv:1605.03204 [gr-qc]} \BibitemShut
  {NoStop}%
\bibitem [{\citenamefont {Zilhao}\ and\ \citenamefont
  {Loffler}(2013)}]{Zilhao:2013hia}%
  \BibitemOpen
  \bibfield  {author} {\bibinfo {author} {\bibfnamefont {M.}~\bibnamefont
  {Zilhao}}\ and\ \bibinfo {author} {\bibfnamefont {F.}~\bibnamefont
  {Loffler}},\ }\bibfield  {booktitle} {\emph {\bibinfo {booktitle}
  {{Proceedings, Spring School on Numerical Relativity and High Energy Physics
  (NR/HEP2): Lisbon, Portugal, March 11-14, 2013}}},\ }\href {\doibase
  10.1142/S0217751X13400149} {\bibfield  {journal} {\bibinfo  {journal} {Int.
  J. Mod. Phys.}\ }\textbf {\bibinfo {volume} {A28}},\ \bibinfo {pages}
  {1340014} (\bibinfo {year} {2013})},\ \Eprint
  {http://arxiv.org/abs/1305.5299} {arXiv:1305.5299 [gr-qc]} \BibitemShut
  {NoStop}%
\bibitem [{\citenamefont {Abbott}\ \emph
  {et~al.}(2016{\natexlab{d}})\citenamefont {Abbott} \emph
  {et~al.}}]{Abbott:2016drs}%
  \BibitemOpen
  \bibfield  {author} {\bibinfo {author} {\bibfnamefont {B.~P.}\ \bibnamefont
  {Abbott}} \emph {et~al.} (\bibinfo {collaboration} {LIGO Scientific
  Collaboration, Virgo Collaboration}),\ }\href {\doibase
  10.3847/0067-0049/227/2/14} {\bibfield  {journal} {\bibinfo  {journal}
  {Astrophys. J. Suppl.}\ }\textbf {\bibinfo {volume} {227}},\ \bibinfo {pages}
  {14} (\bibinfo {year} {2016}{\natexlab{d}})},\ \Eprint
  {http://arxiv.org/abs/1606.03939} {arXiv:1606.03939 [astro-ph.HE]}
  \BibitemShut {NoStop}%
\bibitem [{\citenamefont {Abbott}\ \emph
  {et~al.}(2016{\natexlab{e}})\citenamefont {Abbott} \emph
  {et~al.}}]{Abbott:2016nhf}%
  \BibitemOpen
  \bibfield  {author} {\bibinfo {author} {\bibfnamefont {B.~P.}\ \bibnamefont
  {Abbott}} \emph {et~al.} (\bibinfo {collaboration} {LIGO Scientific
  Collaboration, Virgo Collaboration}),\ }\href {\doibase
  10.3847/2041-8205/833/1/L1} {\bibfield  {journal} {\bibinfo  {journal}
  {Astrophys. J.}\ }\textbf {\bibinfo {volume} {833}},\ \bibinfo {pages} {L1}
  (\bibinfo {year} {2016}{\natexlab{e}})},\ \Eprint
  {http://arxiv.org/abs/1602.03842} {arXiv:1602.03842 [astro-ph.HE]}
  \BibitemShut {NoStop}%
\bibitem [{\citenamefont {Biswas}\ \emph {et~al.}(2009)\citenamefont {Biswas},
  \citenamefont {Brady}, \citenamefont {Creighton},\ and\ \citenamefont
  {Fairhurst}}]{Biswas:2007ni}%
  \BibitemOpen
  \bibfield  {author} {\bibinfo {author} {\bibfnamefont {R.}~\bibnamefont
  {Biswas}}, \bibinfo {author} {\bibfnamefont {P.~R.}\ \bibnamefont {Brady}},
  \bibinfo {author} {\bibfnamefont {J.~D.~E.}\ \bibnamefont {Creighton}}, \
  and\ \bibinfo {author} {\bibfnamefont {S.}~\bibnamefont {Fairhurst}},\ }\href
  {\doibase 10.1088/0264-9381/26/17/175009, 10.1088/0264-9381/30/7/079502}
  {\bibfield  {journal} {\bibinfo  {journal} {Class. Quant. Grav.}\ }\textbf
  {\bibinfo {volume} {26}},\ \bibinfo {pages} {175009} (\bibinfo {year}
  {2009})},\ \bibinfo {note} {[Erratum: Class. Quant. Grav.30,079502(2013)]},\
  \Eprint {http://arxiv.org/abs/0710.0465} {arXiv:0710.0465 [gr-qc]}
  \BibitemShut {NoStop}%
\bibitem [{\citenamefont {Dhurandhar}\ \emph {et~al.}(2017)\citenamefont
  {Dhurandhar}, \citenamefont {Gupta}, \citenamefont {Gadre},\ and\
  \citenamefont {Bose}}]{Dhurandhar:2017aan}%
  \BibitemOpen
  \bibfield  {author} {\bibinfo {author} {\bibfnamefont {S.}~\bibnamefont
  {Dhurandhar}}, \bibinfo {author} {\bibfnamefont {A.}~\bibnamefont {Gupta}},
  \bibinfo {author} {\bibfnamefont {B.}~\bibnamefont {Gadre}}, \ and\ \bibinfo
  {author} {\bibfnamefont {S.}~\bibnamefont {Bose}},\ }\href {\doibase
  10.1103/PhysRevD.96.103018} {\bibfield  {journal} {\bibinfo  {journal} {Phys.
  Rev.}\ }\textbf {\bibinfo {volume} {D96}},\ \bibinfo {pages} {103018}
  (\bibinfo {year} {2017})},\ \Eprint {http://arxiv.org/abs/1708.03605}
  {arXiv:1708.03605 [gr-qc]} \BibitemShut {NoStop}%
\bibitem [{\citenamefont {Abbott}\ \emph
  {et~al.}(2017{\natexlab{e}})\citenamefont {Abbott} \emph
  {et~al.}}]{O1-IMBHB}%
  \BibitemOpen
  \bibfield  {author} {\bibinfo {author} {\bibfnamefont {B.~P.}\ \bibnamefont
  {Abbott}} \emph {et~al.} (\bibinfo {collaboration} {LIGO Scientific,
  Virgo}),\ }\href {\doibase 10.1103/PhysRevD.96.022001} {\bibfield  {journal}
  {\bibinfo  {journal} {Phys. Rev.}\ }\textbf {\bibinfo {volume} {D96}},\
  \bibinfo {pages} {022001} (\bibinfo {year} {2017}{\natexlab{e}})},\ \Eprint
  {http://arxiv.org/abs/1704.04628} {arXiv:1704.04628 [gr-qc]} \BibitemShut
  {NoStop}%
\end{thebibliography}%

\end{document}